\documentclass{aa}  

\usepackage{graphicx}
\usepackage{caption}
\usepackage{txfonts}
\usepackage{xcolor}
\usepackage{hyperref}
\usepackage{float} 
\usepackage{tabularx} 
\usepackage{adjustbox} 
\usepackage{amsmath}
\usepackage{enumitem}
\usepackage{hyperref}
\usepackage{orcidlink}

\hypersetup{
    colorlinks=true,
    citecolor=blue,
    filecolor=blue,      
    urlcolor=blue,
    linkcolor = blue,
    }
    
\newcommand{\nthp}{N$_2$H$^+$\xspace}

\newcommand{\hdos}{H$_2$\xspace}
\newcommand{\ceto}{C$^{18}$O\xspace}
\newcommand{\hfoa}{H41$\alpha$\xspace}
\newcommand{\hh}{\ion{H}{ii}\xspace}

\newcommand{\kms}{\rm km\,s^{-1}}

\newcommand{\degree}{\mbox{$^{\circ}$}\xspace}

\begin{document}

   \title{ALMA-IMF. XXI.: \nthp  kinematics in the G012.80 protocluster}

   \subtitle{Evidence for filament rotation and evolution}
	\titlerunning{N$_2$H$^+$ evidence of filament rotation and evolution in the G012.80 protocluster}
	
   \author{J. Salinas\inst{1,2} \orcidlink{0009-0009-4976-4320} 
          ,
          A. M. Stutz\inst{1,3} \orcidlink{0000-0003-2300-8200}
          ,
          R. H. Álvarez-Gutiérrez\inst{4,1} \orcidlink{0000-0002-9386-8612}
          ,
          N. A. Sandoval-Garrido\inst{1} \orcidlink{0000-0001-9600-2796}
          ,
          F. Louvet\inst{5} \orcidlink{0000-0003-3814-4424}
          ,
          R. Galván-Madrid\inst{2} \orcidlink{0000-0003-1480-4643}
          ,
          F. Motte\inst{5} \orcidlink{0000-0003-1649-8002}
          ,
          M. Armante\inst{6}
          ,
          T. Csengeri\inst{7} \orcidlink{0000-0002-6018-1371}
          ,
          J. Braine\inst{7}
          ,
          A. Ginsburg\inst{8} \orcidlink{0000-0001-6431-9633}
          ,
          M. Valeille-Manet\inst{5,7} \orcidlink{0009-0005-5343-1888}
          ,
          L. Bronfman\inst{9} \orcidlink{0000-0002-9574-8454}
          ,          
          P. Sanhueza\inst{10} \orcidlink{0000-0002-7125-7685}
          ,
          D. Díaz\inst{11} 
          ,
          G. Busquet\inst{12,13,14} \orcidlink{0000-0002-2189-6278}
          ,
          A. Koley\inst{1} \orcidlink{0000-0003-2713-0211}
          ,
          M. Bonfand\inst{15} \orcidlink{0000-0001-6551-6444}
          ,
          M. Fernández-López\inst{16} \orcidlink{0000-0001-5811-0454}
          ,
          A. Gusdorf\inst{17} \orcidlink{0000-0002-0354-1684}
          ,
          N. Castro-Toledo\inst{1} \orcidlink{0009-0005-6363-5104}          
          ,
          R. Veyry\inst{5} \orcidlink{0009-0003-7980-6583}
          ,
          G. Bernal-Mesina\inst{1} \orcidlink{0009-0004-1442-3060} \thanks{Affiliations can be found after the references.}
 }
\authorrunning{J. Salinas et al.}
\institute{}

   \date{Received 07 October 2025 / Accepted 22 May 2026}

  \abstract
  {We aim to characterize kinematic processes in the G012.80 protocluster. We primarily focus on the N$_2$H$^+$(1$-$0) emission to trace the dense and cold gas. Additionally, we use complementary lines such as DCN(3$-$2), H41$\alpha$, C$^{18}$O(2$-$1), and SiO(5$-$4), as well as continuum maps. We perform a N$_2$H$^+$ hyperfine spectral line fitting to analyze multiple velocity components and extract spectral parameters. We estimate velocity gradients, column densities, and line-mass profiles for the two main filaments in G012, which we name R1 and R2. Line-mass profiles follow $\lambda$($\omega$)\,=\,5660\,M$_{\odot}$\,pc$^{-1}$($\omega$/pc)$^{0.30}$ for R1 and $\lambda$($\omega$)\,=\,6943\,M$_{\odot}$\,pc$^{-1}$($\omega$/pc)$^{0.20}$ for R2, which are much larger than those of typical low-mass filaments. R1 and R2 show disparate position-velocity (PV) features. R1 exhibits a transverse velocity gradient of 10.4\,km\,s$^{-1}$\,pc$^{-1}$ and few dense cores. This velocity gradient is interpreted with a simple rotation toy model when combined with line-mass profile, and corresponds to a rotational timescale of $\sim\,0.1$\,Myr. In contrast, R2 exhibits compact velocity structures ($\Delta$V < 2\,km\,s$^{-1}$), likely due to collapse, as evidenced by the presence of a comparatively large number of massive cores and protostellar outflows. R2 is forming prestellar and protostellar cores at a rate of $\sim$ 55.3\,M$_{\odot}$\,Myr$^{-1}$, with an efficiency similar to the Orion Integral Shaped Filament (ISF). The R1 filament, in contrast, lacks protostellar cores and only contains a few prestellar cores, resulting in an estimated star formation rate (SFR) of $\sim$ \,4.2\,M$_{\odot}$\,Myr$^{-1}$, more than an order of magnitude below that of R2. Combining gas kinematics, core incidence, and the line-mass profiles, we suggest that R1 is younger and still rotating, while R2 has evolved to collapse with a higher SFR. G012 thus hosts filaments at different evolutionary stages.}


   \maketitle

\section{Introduction}\label{sec:introduction}

Protoclusters, or embedded star clusters, are gas-dominated regions where stars are actively forming. Unlike clusters where gravity is dominated by the stars themselves, protoclusters are defined by the gravitational influence of the dense gas from which the stars emerge \citep*{stutz2018}. Studying these regions is crucial for understanding the early stages of star cluster formation, particularly the formation of high-mass stars, which significantly impact their surroundings through outflows, radiation, and supernovae, and predominantly form in clusters \citep{motte2018}. These massive stars play a vital role in shaping galaxy evolution and estimating star formation rates across the universe, highlighting the importance of studying protoclusters and their internal assembly processes to understand key astrophysical phenomena.

\begin{figure*}[h!]
  \centering
  \includegraphics[trim=0cm 19cm 0cm 0cm, clip,width=1\linewidth]{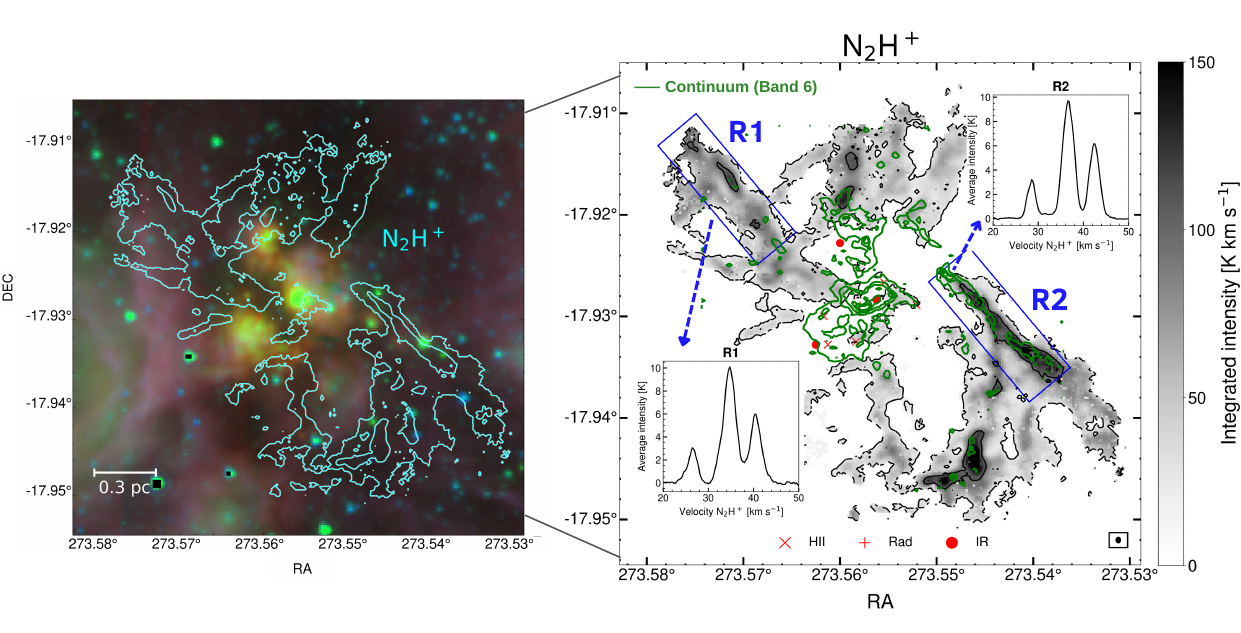} 
  \vspace{-0.8cm}
  \caption{Multi-wavelength view and molecular gas distribution of the G012 protocluster. Left-hand side (l.h.s): Spitzer RGB composite figure of the G012 protocluster at 8~$\mu$m (red), 4.5~$\mu$m (green) and 3.6~$\mu$m (blue). The two cyan contours trace the \nthp integrated intensity emission at 25 and 100\,K\,$\kms$, respectively. Right-hand side (r.h.s): \nthp integrated intensity map with contours at 25 and 100\,K\,$\kms$ corresponding to a $\mathrm{S/N} > 12$, shown in black (same levels as in the left panel). Blue boxes highlight the two main filamentary structures, R1 and R2, with estimated lengths of 0.56\,pc for both. We show the average spectra of both regions in the insets, the black curve represents the data within the black boxes. Red markers represent ionizing regions (X markers), radio sources (+ markers), and infrared sources (dot markers) detected in the region by \citet{haschick1983}. Green contours show continuum emission from ALMA-IMF Band 6 observations \citep[][]{continuum}. The black ellipse in the bottom-right corner represents the beam size of the \nthp data. The integrated intensity map reveals a filamentary and clumpy morphology that traces the dense gas in the region, with a clear absence of emission toward the protocluster center, consistent with the potential chemical destruction of \nthp in the vicinity of the young stellar cluster (see Sec. \ref{sec:nthpdestruction}).}
  \label{fig:SPITZER_N2H+filaments}
\end{figure*}

Molecular gas is often organized in filaments: elongated and dense gas structures that are the birth sites of prestellar and protostellar cores. These cores, embedded within the filaments, represent the sites of star formation and provide a natural link to investigate the kinematic relationship between dense gas and forming protostars. Filaments can be detected at the scale of molecular clouds but also at small protocluster scales (<\,1\,pc); they constitute the immediate gas reservoir from which cores accrete \citep[e.g.,][]{kirk2013,stutz2016,zhou2022,hacar2023,fabien2024}. In cluster-forming environments, filaments are not expected to be static structures. High-mass star formation (HMSF) environments are characterized by deep gravitational potentials, high gas surface densities, stellar feedback, and strong gas motions (such as inflow, turbulence, outflow, among others) all of which can significantly modify the kinematic state of the filamentary gas. Observational signatures encoded in molecular spectra such as longitudinal and transverse velocity gradients \citep[e.g.,][]{gutierrez2021,yiyo2024,nico2024}, velocity coherent substructures \citep[e.g.,][]{gonzalez2019, DCN}, or multiple velocity components \citep[e.g.,][]{csengeri2011a,csengeri2011b} can trace gas accretion flows inside filaments \citep[e.g.,][]{ kong2019, liu2023}, gravitational collapse, rotational motions, or the impact of ionizing feedback and gas ejection \citep[e.g.,][]{galvan2010, roberto2024, towner2024, armante2024}. Distinguishing between these scenarios is particularly important in massive protoclusters, where different theoretical frameworks predict different roles for filamentary gas. For example, global hierarchical collapse models suggest that filaments funnel material toward cluster centers through gravitationally driven inflows \citep[e.g.,][]{enrique2019}, while alternative scenarios emphasize converging turbulent flows \citep[e.g.,][]{padoan2002} or rotationally supported structures as mechanisms shaping the filament dynamics \citep[e.g.,][]{andre2014}. Therefore, studying the kinematics of dense gas in filaments at protocluster scales provides a direct way to test how mass is assembled and redistributed during the formation of stellar clusters.

To study these processes in a massive cluster-forming environment, we focus on the G012.80 protocluster (hereafter G012), one of the 15 nearby massive protoclusters observed by the ALMA-IMF Large Program\footnote{Proposal ID 2017.1.01355.L. ALMA-IMF aims to probe the origin of the initial mass function (IMF) in our Galaxy, providing unprecedented multi-scale data on cores, filaments, and chemically rich regions. This LP focuses on a variety of separate but related topics and techniques, such as the analysis of the different populations of cores in the sample \citep[e.g.,][]{DCN,pouteau2022,pouteau2023,nony2023,fabien2024,motte2024}, search for chemically rich regions using complex organic molecules \citep[e.g.,][]{bonfand2024,armante2024}, study of outflows \citep[e.g.,][]{towner2024,maxime2024}, and also (but not exclusively) gas kinematics in dense filaments \citep[e.g.,][]{yiyo2024,nico2024}. 
} \citep[LP,][]{motte}. Specifically, G012 is an active and massive \citep[M$_{\mathtt{870}\mu m}$ = 1.7\,$\times$\,10$^{3}$\,M$_{\odot}$, see][]{motte} star-forming region centered on the W33 main clump, with a trigonometric parallax distance of 2.4\,kpc \citep{distance} and a local standard of rest velocity of $V_{\mathrm{LSR}}$\,=\,37\,$\kms$ estimated from maser velocities \citep{immer2014}. A central star cluster with spectral types from O7.5 to B1.5 has been reported \citep{distance}, along with evidence of dissociation of complex molecules on small scales \citep[e.g.,][]{immer2014}. Detections of HC$_3$N, CO outflows, class I methanol masers, \hh regions, OB star clusters, and a high abundance of cores indicate that G012 is an optimal testbed for studying HMSF in an evolved protocluster, particularly the mechanisms regulating the distribution and kinematics of dense gas in a feedback-rich environment \citep[e.g.,][]{haschick1983,yu2019,xie2023,armante2024}. To probe the dense gas kinematics in G012, we focus on diazenylium (\nthp; e.g., \citealt{bergin1997,caselli1995,bergin2007,lippok2013,tatematsu2008,busquet2011,gomez2022}). This nitrogen-bearing molecule is characterized by lower depletion levels onto dust grains and was first detected in the interstellar medium by \citet{thaddeus1975}. For the N$_2$H$^+$(1-0) transition, the critical density ranges from 2.0\,$\times$\,10$^4$\,cm$^{-3}$ at 100\,K to 6.1\,$\times$\,10$^4$\,cm$^{-3}$ at 10\,K \citep{shirley2015}, indicating the conditions under which this line becomes an effective tracer. Variations in the observed N$_2$H$^+$ emission can therefore be linked to changes in the physical conditions of the protocluster \citep{tobin2013,tafalla2021,yu2022}.

\begin{table*}[h!]
\caption{Spectral line set-up}
\vspace{-0.5cm}
\begin{center}
\resizebox{\textwidth}{!}{
\begin{tabular}{lccccccccc} 
\hline \hline
Line & Frequency & Velocity resolution & BMAJ$^{a}$ & BMIN$^{b}$ & BPA$^{c}$ & $\mathrm{S/N}^{d}$ & $\rho_{\mathrm{crit}}$ $^{e}$ & E$_{\mathrm{up}}$ $^{f}$ & References \\ 
& [GHz] & [km s$^{-1}$] & [$^{\arcsec}$] & [$^{\arcsec}$] & [\degree] & & [cm$^{-3}$] & [K] & \\ \hline
N$_2$H$^+$(1-0)\,$^{(*)}$ & 93.173 & 0.23 & 2.59 & 2.10 & 89.3 & 12 & 6.1$\times$10$^{4}$ & 4.47 & This paper \\ 
DCN(3-2) & 217.15 & 0.39 & 1.29 & 0.88 & 76.0 & 5 & 1.8$\times$10$^{7}$ & 20.85 & \citet{DCN} \\
H41$\alpha$ & 92.200 & 1.80 & 2.28 & 1.93 & 84.9 & 7 & - & - & \citet{roberto2024} \\ 
C$^{18}$O(2-1) & 219.56 & 0.33 & 1.33 & 0.90 & 77.4 & 5 & 9.9$\times$10$^{3}$ & 5.26 & \citet{atanu2024} \\ 
SiO(5-4) & 217.15 & 0.39 & 1.29 & 0.88 & 77.0 & 5 & 10$^{5}$-10$^{6}$ & 31.25 & \citet{towner2024} \\  
\hline
\end{tabular}
} 
\end{center}
\footnotesize{(a)\,Major axis of the beam.
 (b)\,Minor axis of the beam. 
 (c)\,Beam position angle. 
 (d)\,Signal-to-noise ratio cut applied for the analysis. In all the cases, the $\mathrm{S/N}$ threshold is selected based on the spectral fitting quality (see Sec. \ref{sec:ap_linefitting} for \nthp example).
 (e)\,Critical density of each tracer. Values extracted from \citet{shirley2015,DCN}; and \citet{towner2024}.
 (f)\,Upper energy level extracted from the Cologne Database for Molecular Spectroscopy \citep[CDMS,][]{CDMS}.
 (*) The only tracer that includes total power data.}
\label{tab:SPECTRALINES}
\end{table*}

When the kinematics and dense gas emission traced by molecules like N$_2$H$^+$ are scrutinized in individual protoclusters, multiple evolutionary stages (or generations of stars) emerge within a single region \citep{DCN,pouteau2023,armante2024}. This motivates detailed studies of evolved protoclusters with internal structures at potentially different evolutionary stages. Searching for correlations between the kinematic properties of these dense gas structures \citep*[e.g.,][]{stutz2016,gutierrez2021,xu2023,yiyo2024,reyes2024,nico2024} and their line-mass profiles \citep[e.g.,][]{stutz2018} sets the stage for developing observationally driven physical models of protocluster evolution as stellar mass is assembled. In this context, we focus on the two dominant internal cold dense gas structures in G012 identified with N$_2$H$^+$\,(1-0). These structures have the form of coherent filaments that show evidence for either rotation or collapse. We aim to characterize the dense gas kinematics and key physical parameters in the protocluster, with particular emphasis on these two main filaments. 

This paper is organized as follows. In Sec.~\ref{sec:data} we detail the ALMA-IMF dataset and core catalogs for the G012 protocluster. In Sec.~\ref{sec:analisis} we analyze the \nthp moment maps, PV diagrams, column density, and \nthp core velocities in the region. We perform a detailed analysis of the two dominant cold dense gas filamentary structures in Sec.~\ref{sec:filamentanalisis}, focusing on average-velocity gradients and line-mass profiles. In Sec.~\ref{sec:rotationR1}, we apply a rotation toy model to the observed dense gas kinematics in the R1 filament. In Sec.~\ref{sec:dis} we discuss potential evolutionary scenarios by contrasting the kinematics and other observables (e.g., core incidence). Finally, we summarize our main conclusions in Sec.~\ref{sec:conclusion}.

\section{Data}\label{sec:data}

We use observations from the Atacama Large Millimeter Array~(ALMA) telescope provided by the ALMA-IMF LP. Line parameters for \nthp and additional tracers are summarized in Table~\ref{tab:SPECTRALINES}. 

\subsection{ALMA-IMF datacubes}
We employ $\mathrm{N_2H^+}$($J$=1$-$0) observations at a frequency of 93.1734 GHz. We follow a similar data reduction procedure as \citet{yiyo2024} and \citet{nico2024}. \nthp observations were cleaned using the ALMA-IMF imaging pipeline\footnote{\url{https://github.com/ALMA-IMF/reduction}} \citep{DCN} and the version 5.6.0 of The Common Astronomy Software Applications package \citep[CASA,][]{casateam}. We use the \texttt{imcontsub} task to subtract the continuum emission from the N$_2$H$^+$ line emission. To estimate the continuum emission we consider only emission-free channels, in the range of 12 to 20\,$\kms$, where we apply a linear fit using \texttt{fitorder} = 0. To recover the cloud emission at all available scales, we combine observations of the 7m and 12m arrays with total power data (TP). For this, we use the \texttt{feather} task. This combination also allows us to recover the missing flux, visible as negative bowls in the interferometric spectral data produced by the lack of zero-spacing. We obtain a fully integrated, multi-scale emission dataset. The resulting \nthp data cube contains a total of 217 velocity channels over a range from 12\,$\kms$ to 61\,$\kms$, with a spectral resolution of 0.23\,$\kms$ and a final beam size of $\sim$ 2.3$^{\arcsec}$ (see Table \ref{tab:SPECTRALINES}), corresponding to $\sim$5.5\,kau at the source distance. In Fig.~\ref{fig:SPITZER_N2H+filaments} we display Spitzer observations at 8~$\mu$m (red), 4.5~$\mu$m (green) and 3.6~$\mu$m (blue) along with an N$_2$H$^+$ integrated intensity contours at 25 and 100\,K\,$\kms$ (left panel) and \nthp integrated intensity map at signal-to-noise ratio ($\mathrm{S/N}$)\,>\,12 (right panel, see Appendix \ref{sec:ap_linefitting} for the $\mathrm{S/N}$ selection criteria).

In addition to the \nthp data, we use complementary spectral lines previously analyzed in ALMA-IMF studies. These include the \hfoa recombination line \citep{roberto2024}, which traces \hh regions associated with ionizing sources in the protocluster center; DCN\,($J$=2$-$1; \citealt{DCN}), which traces more compact and warmer emission compared to \nthp; SiO\,($J$=5$-$4; \citealt{towner2024}), used to analyze potential outflow features; and \ceto\,($J$=2$-$1; \citealt{atanu2024}), which traces compact emission in the central region of the protocluster as well as more extended and fainter emission in the surroundings.

\subsection{ALMA-IMF dense core catalogs} 

\citet{fabien2024} constructed the first core catalog of the full ALMA-IMF sample based on the 1.3\,mm and 3\,mm continuum emission. They used the \textit{Getsf} toolkit \citep{getsf} to identify compact continuum sources. In G012, they detected 57 dense cores with estimated masses of 0.6-6.2\,M$_{\odot}$ and temperatures of 25-100\,K. As follow-up, \citet{DCN} used the DCN molecular line emission to study the core population from \citet{fabien2024}. For G012, they were able to estimate spectral parameters of 38 cores, including their line-of-sight velocities. Deeper, \citet{armante2024} constructed a catalog of G012 cores complementing dust continuum emission (at both 1.3\,mm and 3\,mm) with molecular emission from $^{12}$CO\,(2-1), SiO\,(5-4), CH$_3$OCHO, and CH$_3$CN spectral lines. They classified the cores as prestellar or protostellar, finding a total of 94 detections. Additionally, \citet{motte2024} provided new mass, temperature, and luminosity estimates for cores in the remaining ALMA-IMF protoclusters using radiative transfer modeling. For G012, they found protostellar and prestellar cores distributed over mass ranges of 0.3-9.5 M$_{\odot}$. 

To complement these data and provide a more complete characterization of G012, in our analyses we make use of all these core catalogs. In particular, in the Appendix\,\ref{sec:cores} we estimate new velocities for 48 cores from these previous catalogs.

\begin{table*}[h!]
    \caption{Global parameters of the R1 and R2 filaments} 
    \vspace{-0.5cm}
    \begin{center}
    \resizebox{\textwidth}{!}{
    \begin{tabular}{l     c     c     c     c     c     c     c     c     c     c} \hline \hline
        Region & Length$^{a}$ & Width$^{a}$  & <V>$^{b}$ & <$\sigma$>$^{b}$ & <$\tau$>$^{b}$ & <T$_{\mathrm{ex}}$>$^{b}$ & M$_{f}^{c}$ & M$_f$/M$_{tot}^{d}$ & N(H$_2$)$^{e}$ & N(\nthp)$^{f}$ \\ 
         & pc & pc & km/s & km/s &  & K & 10$^{3}$\,M$_{\odot}$ & & 1e$^{26}$ cm$^{-2}$& 1e$^{16}$ cm$^{-2}$\\ \hline
        R1 & 0.56 & 0.22 & 34.4 $\pm$ 0.05 & 0.81 $\pm$ 0.05 & 1.96 $\pm$ 1.4 & 41.2 $\pm$ 7.6 & 1.78 & 0.12 & 0.56 & 6.75\\ 
        R2 & 0.56 & 0.22 & 36.8 $\pm$ 0.07 & 0.75 $\pm$ 0.06 & 2.49 $\pm$ 1.8 & 31.5 $\pm$ 4.7 & 2.53 & 0.17 & 1.27 & 8.26\\ \hline
    \end{tabular}
    }
    \end{center}
\footnotesize{(a) Length and width of the boxes enclosing the main filaments R1 and R2 in Fig.~\ref{fig:SPITZER_N2H+filaments} (b) Mean of the fitted parameters in the dominant velocity component of each filament. (c) Total H$_2$ filament mass determined by N(\nthp) column density and relative abundance (see Sec. \ref{sec:xfact}) (d) Fraction of the filament mass respect to the total protocluster mass, where for the G012 protocluster M$_{tot}$~=~14.8~$\times$~10$^3$~M$_{\odot}$ (see Sec. \ref{sec:dens}). (e) Total column density inside the filaments extracted from \citet{pierre2024} maps. (f) \nthp column density inside the filaments.}
\label{tab:fil}
\end{table*}

\section{Analysis of the \nthp data cube}\label{sec:analisis}

By modeling the hyperfine structure of the N$_2$H$^+$ (as described in Appendix \ref{sec:ap_linefitting}), we derive key kinematic parameters and simultaneously identify multiple velocity components in the \nthp spectra. We adopt a similar fitting procedure as \citet{nico2024} to model the \nthp emission. We use the hyperfine line structure model provided by version V.1.0.1 of \texttt{PySpecKit} spectroscopic analysis toolkit \citep{pyspec}. The vast majority of spectra that we inspected are well characterized by either one or two velocity components (see Appendix~\ref{sec:ap_linefitting}). Hence, we applied a fitting model with up to two components, using parameters such as excitation temperature (T${\mathrm{ex}}$), optical depth ($\tau$), velocity centroid (V${\mathrm{c}}$), and velocity dispersion ($\sigma$). The model fitting results show that G012 is characterized by two main velocity structures separated by $\sim$2.5 $\kms$, which we define as the first velocity component (FVC) and the second velocity component (SVC). We adopt 35.6~$\kms$ as the reference velocity (see Appendix~\ref{sec:ap_linefitting}) to separate them: FVC refers to spectra with velocities below this value, while SVC refers to those above it. Using the FVC and SVC structures, we further explore the kinematics of G012. First, we review moment maps that provide a detailed look at the velocity centroid and dispersion patterns across the region, and of the particular filamentary structures R1 and R2 (see Fig.~\ref{fig:SPITZER_N2H+filaments}). We then construct PV diagrams of the entire region to capture the velocity gradients and dynamical behavior of the protocluster. We also estimate column densities and masses in G012 in order to set constraints on the gas mass and density profiles in the region.

\begin{figure*}[h!]
    \centering
    \includegraphics[trim=0cm 5cm 0cm 0cm, clip,width=0.8\linewidth]{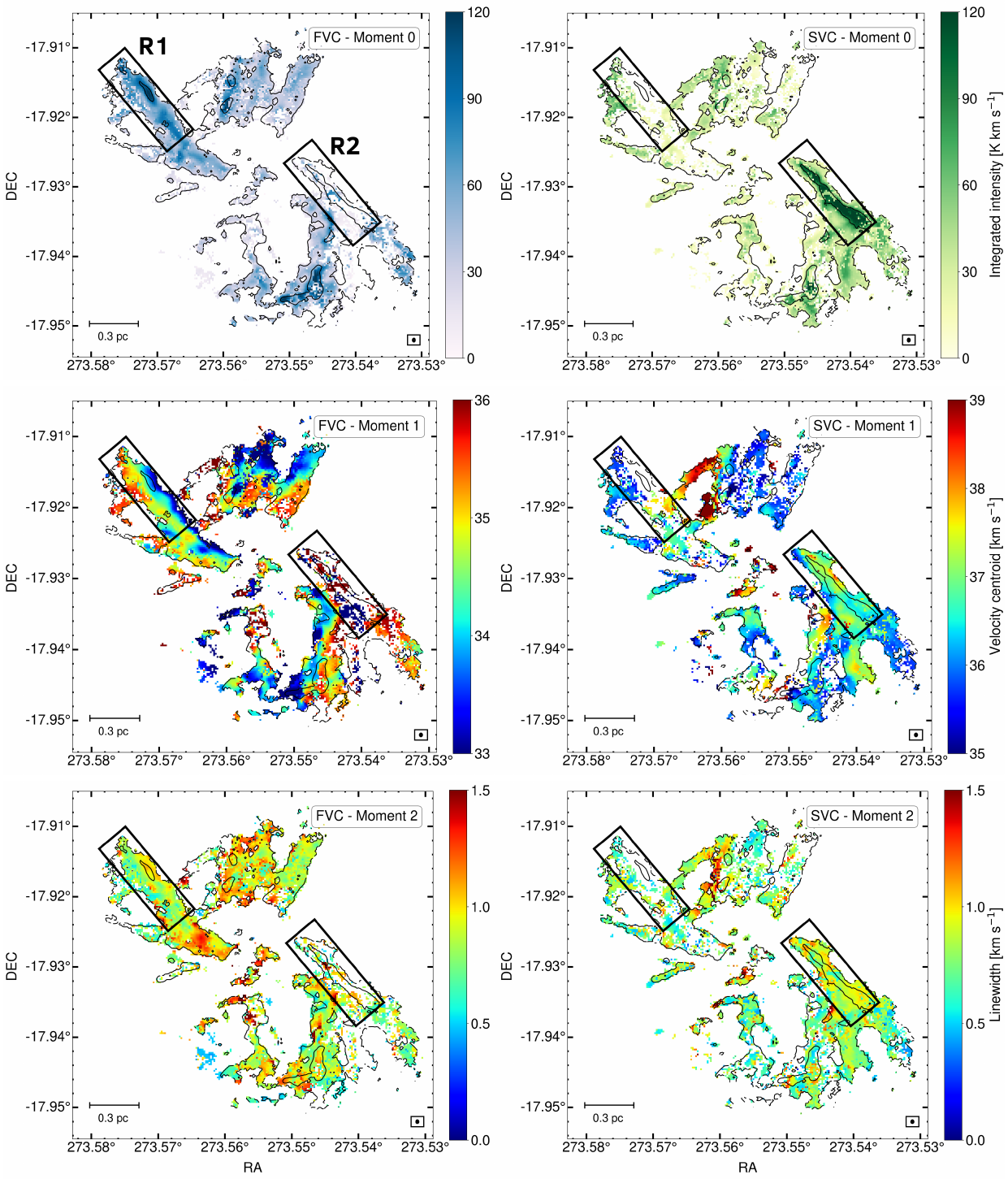}
    \vspace{-0.3cm}
    \caption{N$_2$H$^+$(1$-$0) integrated intensity (upper panels), velocity centroid (middle panels), and velocity dispersion (bottom panels) of the FVC (left panels) and SVC (right panels). Black contours trace the \nthp integrated intensity emission at 25 and 100\,K\,$\kms$, respectively. Black boxes in all panels display the spatial location of the R1 and R2 filaments. The black ellipse at the bottom-right corner represents the beam size of the \nthp data.}
    \label{fig:structures}
\end{figure*}

\subsection{Moment maps}\label{sec:moments}

In Fig.~\ref{fig:structures}, we display the \nthp integrated intensity (moment zero map, upper panels), velocity center (moment one map, middle panels), and velocity dispersion (moment two map, bottom panels) maps of the FVC and SVC. In the \nthp integrated intensity map, G012 is characterized by the presence of a filamentary distribution in the plane-of-the-sky (POS). Specifically, we highlight the two prominent filaments with blue boxes (see also Fig.~\ref{fig:SPITZER_N2H+filaments}), which are the focus of this study, hereafter called R1 and R2. These structures were identified based on the visual inspection of prominent and kinematically well-separated filamentary structures, where the kinematic patterns described below are more clearly distinguished. However, to complement this more visual inspection, we also applied a more automated method, as described in Appendix\,\ref{sec:ap_filfinder}.

The R1 filament is largely dominated by the FVC (see Fig.~\ref{fig:structures}, left panel), which displays one pronounced velocity gradient across the filament. The R1 mean centroid velocity and line width are $\langle V \rangle$\,=\,34.4\,$\kms$ and $\langle \sigma \rangle$\,=\,0.81\,$\kms$, respectively. In contrast, the R2 filament is dominated by the SVC (see Fig.~\ref{fig:structures}, right panel), which exhibits more homogeneous velocity structures. The mean centroid velocity and line width for R2 are $\langle V \rangle$\,=\,36.8\,$\kms$ and $\langle \sigma \rangle$\,=\,0.75\,$\kms$, respectively (see Tab.~\ref{tab:fil} for global parameters of the main filaments). Additionally, at the center of G012 we observe an absence of \nthp emission that may be related to the destruction of the molecule in the region (see Sec.~\ref{sec:dis}). To analyze this absence, we review the integrated intensity maps of the fitted complementary tracers in Fig.~\ref{fig:TRACERS}. We observe that the \ceto emission (upper-left panel) traces more extended gas than \nthp and is primarily distributed around the \hfoa bubbles (bottom-left panel). In addition, we identify high \ceto integrated intensity structures in the surroundings of the R1 and R2 filaments. SiO integrated intensity reveals elongated features in the R2 filament (upper-right panel). Specifically, one perpendicular outflow feature is observed at the top of R2 related to a previous detected hot core \citep{armante2024}. Furthermore, DCN, which is associated with increased star formation, traces some of the densest parts of the filamentary structures in R1 and R2 (bottom-right panel).

\subsection{\nthp PV diagrams}\label{sec:pv}

\begin{figure*}[h!]
    \centering
    \includegraphics[trim=0cm 10cm 0cm 0cm, clip,width=0.95\linewidth]{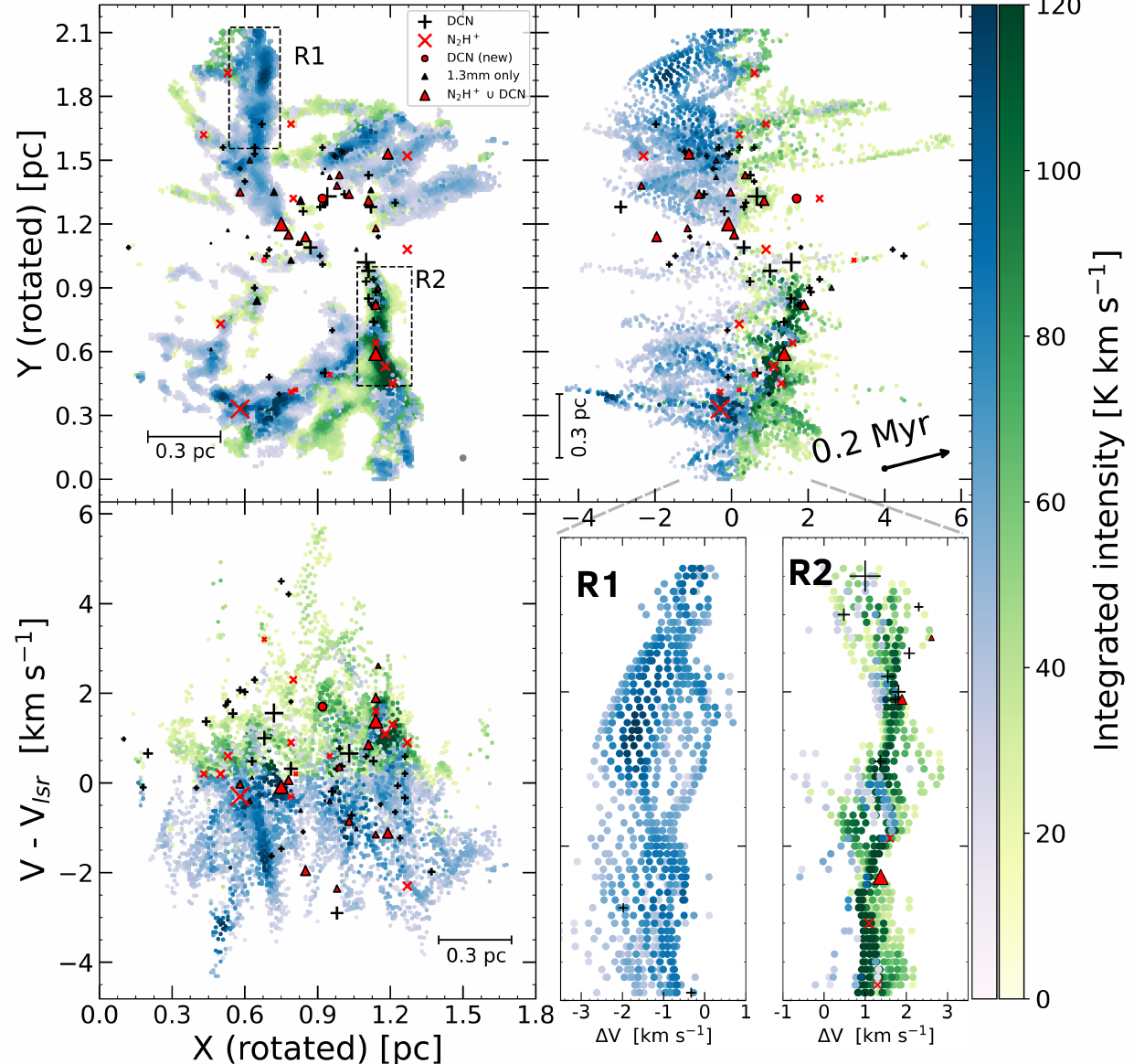}
    \vspace{-0.2cm}
    \caption{Position-position and position-velocity diagrams of the \nthp observations. Upper-left: G012 integrated intensity of FVC (blue shades) and SVC (green shades). We display the DCN cores from \citet{DCN} catalog with black ``+'' symbols. The \citet{armante2024} core catalog is divided into four categories: cores detected only with \nthp (red ``$\times$'' symbols), cores detected only with DCN data (red circles), cores detected with \nthp and DCN (red triangles), and cores neither detected in both \nthp nor DCN (black triangles). Symbol sizes are proportional to the estimated core mass. The black ellipse in the bottom right corner represents the beam size of \nthp data. Upper-right: PV diagram along the y-axis; the velocity axis is subtracted from the \nthp systemic $V_{\mathrm{LSR}}$ of the protocluster (35.5\,$\kms$). The black arrow illustrates a slope of 5\,pc\,($\kms$)$^{-1}$, which corresponds to a timescale of $\sim$\,0.2\,Myr. That is, the timescales that approximately correspond to some of the extended structures in this PV diagram. At the top of R1 we observe a wrapped (or ``double helix'') type velocity field with spreads of more than $\sim$\,3\,$\kms$. Meanwhile, R2 appears very compact in velocity along its extent, with small-scale spatial ``wiggles''. Bottom-left: PV diagram in the perpendicular direction compared to the upper-right panel. Here we observe the emergence, albeit somewhat hidden in the overall velocity field of an approximately uniform and extended gradient in R1 apparent near $\Delta$X~$\sim$~0.60 to 0.75\,pc, $\Delta$V~$\sim$~-2 to 0\,$\kms$. Meanwhile R2 appears as the compact ``blob'' of cores (triangles and x-symbols) on the r.h.s.\ of the panel, characterized by the absence of an obvious gradient in position and velocity. Bottom-right: Zoomed PV diagrams of the main filaments R1 (l.h.s.) and R2 (r.h.s.), enclosed in the black boxes of the upper left panel. R1 presents a wrapping ``double-helix'' signature that is most obvious toward the top of the diagram and which is dominated by the FVC velocities. R2 exhibits comparatively compact velocity variations ($\Delta V_{\mathrm{max}}~\sim~1.5$~km~s$^{-1}$) along the filament, and contains a high number of massive cores (1 M$_{\odot}$ - 3 M$_{\odot}$).}
    \label{fig:pv_sub}
\end{figure*}

\begin{figure*}[h!]
    \centering
    \includegraphics[trim=0cm 20cm 0cm 0cm, clip,width=0.9\linewidth]{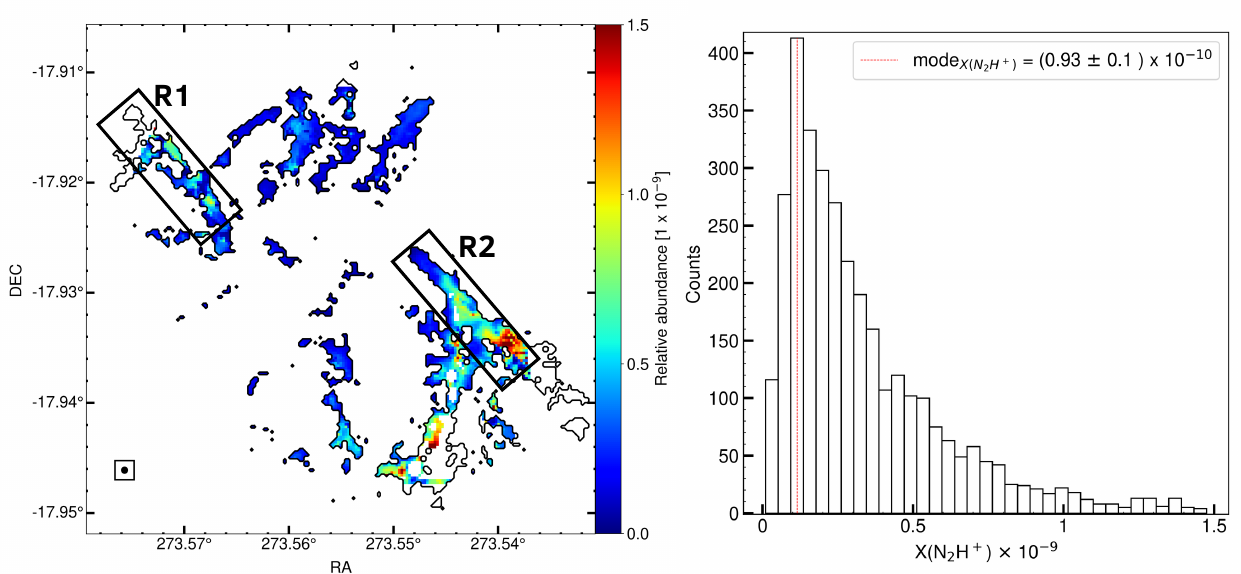}
    \vspace{-0.2cm}
    \caption{Left panel: Relative abundance map in G012. The black contour highlights the mask of $\frac{\tau}{e(\tau)}$\,>\,2 applied to the \nthp column density (see Sec.~\ref{sec:xfact}); most of the pixels removed by this mask do not affect the main filaments significantly. The areas lacking data (top of R1 and bottom of R2) are due to the H$_2$ column density map lack of coverage in the filaments. We find a representative relative abundance value of 0.93\,$\times$\,10$^{-10}$. The black circle in the bottom-left corner represents the beam size of the H$_2$ map. Right panel: Relative abundance histogram of the values inside the black contour in left panel. The red-dashed line represents the mode of the distribution. In the upper-right corner, the values of the mode and standard deviation are shown, which we consider as the representative value and error of the sample.}
    \label{fig:ap_xfactmap}
\end{figure*}

We use the technique developed in \citet{gonzalez2019} to construct the \nthp intensity-weighted position velocity diagrams \citep[e.g.,][]{gutierrez2021,yiyo2024,nico2024}. To obtain better defined structures in PV space, we spatially rotate the \nthp data cubes (45\degree respect to the protocluster center) by vertically aligning the most predominant filaments. In Fig.~\ref{fig:pv_sub} we show PV diagrams of the first velocity component (blue shades) and the second velocity component (green shades) structures. The upper-left panel shows the integrated intensity map after spatial alignment. In the upper-right and bottom-left panels we display the PV diagrams perpendicular and parallel to the region, respectively. Additionally, we include the positions of previous cores detected in G012 \citep[][]{DCN,fabien2024,armante2024,motte2024}. For the PV diagrams, we show a representative {\renewcommand{\thefootnote}{\alph{footnote}}
\setcounter{footnote}{0} timescale of 0.2 Myr\footnote{We consider the timescale $\propto \left( \frac{\Delta v}{\Delta p}\right)$ [Myr], neglecting projection effects and assuming a linear velocity gradient, where $v$ is the velocity axis and $p$ the position axis in the PV.}} related to the PV structures more spread in velocity ($\Delta$V~>~3\,km\,s$^{-1}$). In addition to the main features mentioned above, we also highlight an elongated structure with high integrated intensity in the bottom-left diagram, located at velocity range ($\Delta$V) of approximately -2 to 0\,$\kms$. This structure is spatially related to the R1 filament (see Sec.~\ref{sec:filamentanalisis}). The key features in the PV diagrams are:

\begin{enumerate}
	\item R2 reveals a set of characteristic twisting features (Fig.~\ref{fig:pv_sub}, bottom-right panel) which are confined to a narrow velocity range (<~2~$\kms$), in contrast to the broader velocity spread observed in R1.
	\item The PV features are more spread out in velocity ($\Delta$V~>~3\,km\,s$^{-1}$, see upper-right panel in Fig.~\ref{fig:pv_sub}) in regions near \hfoa and SiO emission (see Fig.~\ref{fig:TRACERS}), indicating short associated timescales. This potentially highlights the effect of stellar feedback close to the center of G012.
	\item The R1 and R2 filaments reveal the most predominant and intricate PV structures .
	\item The R1 filament exhibits a ``double-helix'' feature (Fig.~\ref{fig:pv_sub}, bottom-right panel), spanning approximately $\sim$~3~$\kms$ in velocity. This kind of feature has been associated with filament rotation in other regions \citep[e.g.,][]{gutierrez2021}, a point that we address in Sec.~\ref{sec:rotationR1}.
\end{enumerate}

\subsection{\nthp column density}\label{sec:dens}

Column density and mass in protoclusters provide insights into the amount of material available for star formation. In order to estimate the \nthp column density, we use the T$_{\mathrm{ex}}$, $\tau$, and $\sigma$ \texttt{PySpecKit} output parameters (see Appendix \ref{sec:ap_linefitting}). We apply the column density approximation outlined in \citet{caselli2002a} and \citet{readelli2019}, which is described as follows:

\begin{equation}
    \mathrm{N}(\mathrm{N}_2\mathrm{H}^+) = 
\frac{4 \pi^{3/2} \, \nu^3 \, Q \, \sigma \, \tau \,
\exp\!\left(\tfrac{E_u}{k_\mathrm{B} T_\mathrm{ex}}\right)}
{\sqrt{\ln 2} \, c^3 \, A \, g \,
\left(\exp\!\left(\tfrac{h \nu}{k_\mathrm{B} T_\mathrm{ex}}\right) - 1\right)}.
    \label{eq:cdens}
\end{equation}
Here, $\nu$ represents frequency of the N$_2$H$^+$(1$-$0) emission, g\,=\,3 is the statistical weight (degeneracy) of the upper energy level of the transition level, c is the light speed, A\,=\,3.6e$^{-5}$ s$^{-1}$ is the Einstein coefficient, $\text{E}_\text{u}$\,=\,4.47\,K corresponds to the upper energy level of the transition, Q is the rotational partition function of \nthp, $\text{k}_\text{B}$ is the Boltzmann constant, and $h$ is the Planck constant \citep{shirley2015,readelli2019}.

To mitigate potential biases introduced by poorly defined fitting parameters (see Appendix \ref{sec:ap_linefitting}) and prevent anomalies in the column density and relative abundance estimates, we excluded all pixels influenced by systematic biases in $\tau$ and T$_{\mathrm{ex}}$. Specifically, we removed pixels where T$_{\mathrm{ex}}$\,=\,150\,K (the fitting upper limit) and $\tau$\,<\,0.2, which resulted in discarding 30$\%$ of the affected pixels (see Appendix \ref{sec:ap_relative}). We consider the final column density map as the sum of the FVC and the SVC column densities. Overall, the R1 column density distribution appear more interconnected, compact, and filamentary. Meanwhile, the R2 column density values are distributed more homogeneously, taking high values of around 1\,$\times$\,10$^{14}$\,cm$^{-2}$ in most of the region. We convert \nthp column density to mass, obtaining a respective \nthp mass of 1.72\,$\times$\,10$^{-5}$\,M$_{\odot}$ in the entire protocluster. The two main filaments, R1 and R2, contribute approximately 12$\%$ and 17$\%$ of the \nthp protocluster mass, respectively. Moreover, when considering the extended emission associated with these filaments (see Appendix\,\ref{sec:ap_filfinder}), their total contribution accounts for about 60\,$\%$ of the total \nthp mass, further emphasizing their importance in both the kinematics and mass distribution of the region.

\subsection{Relative abundance}\label{sec:xfact}
In G012, previous estimates of the H$_2$ column density map \citep{pierre2024} do not provide the full spatial coverage achieved by the \nthp map. Nevertheless, despite these limitations, particularly the incomplete coverage of the main filaments in the region, the map from \citet{pierre2024} remains the most suitable tool currently available to estimate the relative abundance in the region. We use the \nthp column density to estimate the total mass (M$_\mathrm{tot}$) in the protocluster as a whole, and in areas where the H$_2$ map lacks complete spatial coverage. First, we calculate the relative abundance between both \nthp and H$_2$ column density maps (see Eq.~\ref{eq:xfact}) where we have coverage in both. To estimate this ratio, we reproject the \nthp image to match the pixel scale of the H$_2$ data, which is slightly larger (with a pixel scale of 0.83$^{\arcsec}$). Subsequently we calculate the relative abundance map by taking the pixel-to-pixel ratio of the two maps: 
\begin{equation}
    X(N_2H^+) = \frac{N(N_2H^+)}{N(H_2)} 
    \label{eq:xfact}
\end{equation}
Here, N(\nthp) is the \nthp column density map described above, and N(H$_2$) is the H$_2$ column density map from \citet{pierre2024}. For the N(H$_2$) map, we consider a $\mathrm{S/N}$\,>\,3 based on the ratio between the column density and the associated errors. In Fig.\,\ref{fig:ap_xfactmap} (left panel), we present the resulting relative abundance map. We identify a clear trend in which the  relative abundances values decrease toward regions associated with ionization (see Fig.\,\ref{fig:dcn_pv} left panel, for a comparison between the \nthp emision and the \hfoa distribution). This behavior likely reflects the impact of stellar feedback on the molecular gas distribution, a point that we further discuss in Sec.\,\ref{sec:dis}.

\begin{figure*}[h]
    \centering
    \includegraphics[trim=0cm 19.5cm 0cm 0cm, clip,width=0.9\linewidth]{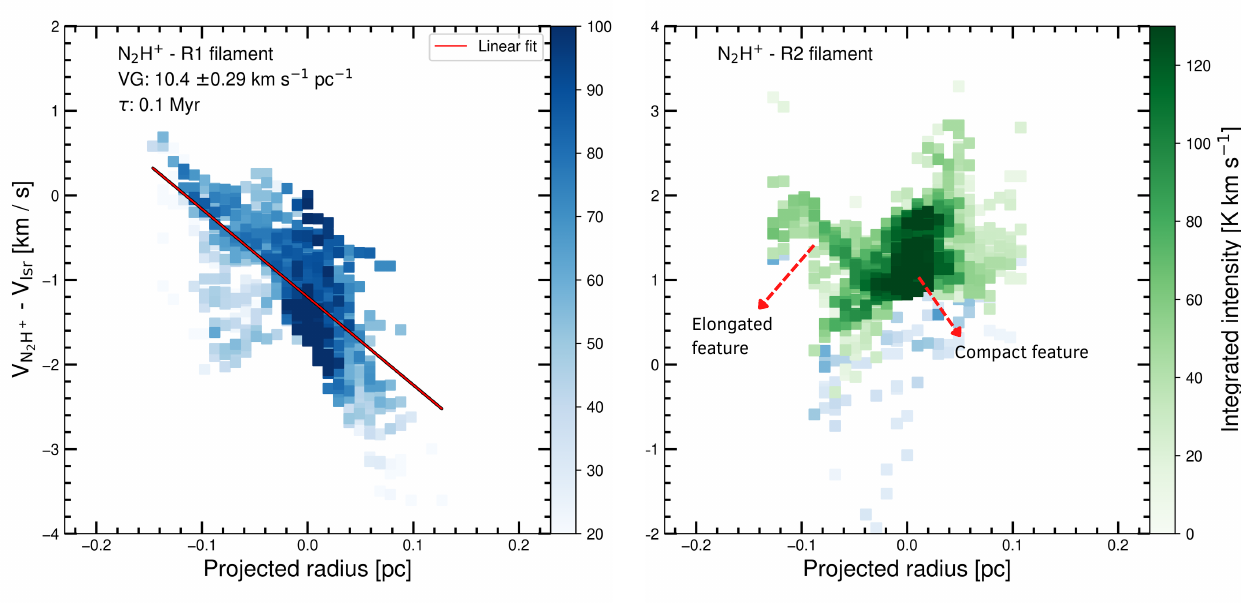}
    \vspace{-0.5cm}
    \caption{Average velocity gradients perpendicular to R1 (l.h.s.) \& R2 (r.h.s.) filaments. In both diagrams, the $\Delta$V and $\Delta$r axes have the same range, allowing for direct comparison of slopes between panels. In R1 the red line represents the linear fit weighted by the integrated intensity points of the velocity gradient (VG). The associated VG timescale ($\tau$) estimation is displayed in the upper left corner. In opposition, R2 lacks a clear velocity gradient structure as in R1 and is instead comparatively compact in its velocity distribution. In addition, we identify two different structures that spatially correspond to different regions in the R2 filament. The most compact and central structure (compact feature) is related to the densest region of the filament. The most elongated feature is spatially related to the eastern edge of the R2 filament, characterized by the presence of elongated SiO emission. In the diagram, the lower intensity and scattered points represent regions surrounding the filament.}
\label{fig:gradients}
\end{figure*}

We determine representative N$_2$H$^+$ relative abundance values using the mode of its distribution, a choice motivated by its stability under different binning and masking conditions (see Appendix~\ref{sec:ap_relative}). We obtain a relative abundance of X(N$_2$H$^+$) = (0.93 $\pm$ 0.10)\,$\times$\,10$^{-10}$ (see Figure~\ref{fig:ap_xfactmap}, right panel), consistent with previous determinations \citep{caselli2002c,nico2024}. This representative value is then applied to derive total H$_2$ mass. Since these estimates will underpin the line-mass profiles presented in Sec.~\ref{sec:linemass}, we first validate our choice of mass map by calculating the line-mass profile in a section of the R2 filament covered by both the N$_2$H$^+$ and \citet{pierre2024} column density maps (see Appendix \ref{sec:ap_massval}). We find variations of less than 28$\%$ between the two methods, indicating that the use of N$_2$H$^+$ to re-estimate the total H$_2$ mass, and consequently the line-mass profiles discussed in Sec.~\ref{sec:linemass}, is robust.

\section{R1 and R2 characterization}\label{sec:filamentanalisis}

The R1 and R2 filaments are the predominant \nthp integrated intensity structures in the G012 protocluster. These regions present different velocity structures in PV diagrams (see Sec.~\ref{sec:analisis}), suggesting the presence of two different star forming environments taking place in the same protocluster. In this section we focus on the characterization of these two filaments, aiming to set constraints on their star forming stage.

 \subsection{Average filament velocity gradients}\label{sec:gradients}

As shown in Fig.~\ref{fig:pv_sub} (bottom-right panel), the R1 filament is associated with a potential signature of rotation, while R2 presents a relatively compact velocity distribution. Here we analyze the velocity gradients perpendicular to both filaments through average-velocity gradients estimations. We adapt of the method described in \citet{gutierrez2021} for filament rotation analysis and divide the procedure into the following steps:

\begin{enumerate}
       \item We determine the total integrated intensity in each region (i.e., the moment 0 of the FVC and SVC, see Sec.~\ref{sec:analisis}) and calculate peak integrated intensity values along the long axis of both filaments in order to determine a representative filament ridgeline (see Appendix\,\ref{sec:ap_datalign}).
	\item For each slice along the Y axis we straighten both filaments by centering the peak integrated intensity from the ridgeline to a fixed arbitrary position. Then, each point of the integrated intensity, velocity centroid, and velocity dispersion maps has a projected radius to the center of the filament.

	\item  We generate ``ridgeline-average'' PV diagrams as follows. For each point, at each radius from the ridgeline, we plot the radial velocities as shown in Fig.~\ref{fig:gradients}. Here the color of each point indicates their integrated intensity. These velocity vs. radius diagrams therefore capture PV structures integrated over the length of the filament. If a given filament has a prominent velocity gradient approximately perpendicular to the ridgeline, these diagrams will reveal these structures. Since the projected radius is measured perpendicular to the filament ridgeline, a coherent velocity gradient in these velocity vs. projected radius diagrams is naturally interpreted as rotation around the filament axis. This is the case for R1, while for R2 no such equivalently prominent structure is observed. 
	
	\item We apply a linear fit to the velocity gradient in the R1 filament, using the integrated intensity as a statistical weight.
\end{enumerate}

\begin{table*}[h!]
\caption{Line-mass profiles, volume density, gravitational potential, and acceleration distributions}
	\vspace{-0.5cm}
    \begin{center}
    \begin{tabular}{l c c c c c c c} \hline \hline 
        Region & $\zeta^{a}$ & $\beta^{b}$ & $\psi^{c}$ & $\xi^{d}$ & $\gamma^{e}$ & Projected Length$^{f}$ & Total Gas Mass$^{g}$ \\ 
        
         & M$_{\odot}$pc$^{-1}$ & M$_{\odot}$pc$^{-3}$ & km$^2$s$^{-2}$ & km$^2$s$^{-2}$pc$^{-1}$ &  & pc & M$_{\odot}$ \\
         \hline 
         \\
         
        G012.80-R1 & 5660 & 211.3 & 123.1 & 37.4 & 0.30 & 0.56 & 1780 \\ 
        
        G012.80-R2 & 6943 & 181.3 & 269.3 & 51.3 & 0.20 & 0.56 & 2532 \\ \\
        \hline 
        \\
        
        L1482-S$^{h}$   & 205  & 9.60  & 1.27 & 0.81 & 0.64 & 0.90 & 380  \\ 
        
        ONC$^{h}$        & 866  & 25.9  & 27.6 & 6.40 & 0.23 & 0.50 & 1300 \\ 
        
        Orion ISF$^{h}$  & 385  & 16.5  & 6.30 & 2.40 & 0.38 & 7.30 & 6200 \\ \\
        \hline 
        \\
        
        G351.77$^{h}$    & 1660 & 78.7  & 13.5 & 8.38 & 0.62 & 8.60 & 10200\\  \\
        \hline
        \\
        
        R1-EXT$^{i}$ & 5764 & 256.6 & 81.6 & 33.6 & 0.41 & 0.93 & 2450 \\ 
        
        R2-EXT$^{i}$ & 8000 & 384.6 & 83.6 & 41.6 & 0.49 & 1.0 & 5000 \\ \\
        \hline

    \end{tabular}
    \end{center}
    \footnotesize{Normalization constants for: (a) the M/L profile in Eq. \ref{eq:linemassprofile}, (b) the volume density in Eq. \ref{eq:volumedensity}, (c) the gas gravitational potential in Eq. \ref{eq:gravitationalpot}, and (d) the gravitational acceleration in Eq. \ref{eq:gravitationalacc}. (e) Power-law index in the M/L profile. (f) Filament length for the M/L estimates (see Sec. \ref{sec:linemass}). (g) Total H$_2$ mass in the filaments, estimated using the \nthp relative abundance (see Sec. \ref{sec:xfact}). (h) Regions described in Sec. \ref{sec:linemass} and shown in Fig. \ref{fig:enclosedmass}. (i) Metrics inside extended R1 and R2 filaments, see Appendix \ref{sec:ap_filfinder}.}
\label{tab:metrics}
\end{table*}

In Fig.~\ref{fig:gradients} we show the R1 (l.h.s.) and R2 (r.h.s.) ridgeline-averaged PV diagram described above. In this representation, a linear and elongated structure reflects a systematic change of velocity with projected radius, as expected from rotation around the filament axis. The global R1 velocity gradient (VG) has a magnitude of VG\,$\sim$\,10.4\,km\,s$^{-1}$\,pc$^{-1}$ (with an associated statistical fitting error of $\pm$ 0.29\,km\,s$^{-1}$\,pc$^{-1}$). In R2, in sharp contrast to R1, we do not observe an equivalent global average velocity gradient. In contrast, R2 is dominated by a compact velocity structure, clumping at velocities around $\sim\,$1\,km\,s$^{-1}$ in the center of the diagram. This compact feature traces the central and densest part of the filament. We also observe a more spread out structure in projected radius and velocity extending up and to the left in the diagram and ending near -~0.05\,pc, 0.75\,$\kms$ (elongated feature). This structure is associated with the $\sim$ eastern edge of the R2 filament, and has associated elongated SiO emission likely tracing protostelar outflows and shocks \citep{towner2024}.

\subsection{Line-mass profiles and associated 3D model quantities}\label{sec:linemass}

\begin{figure}[h!]
    \includegraphics[width=1\linewidth]{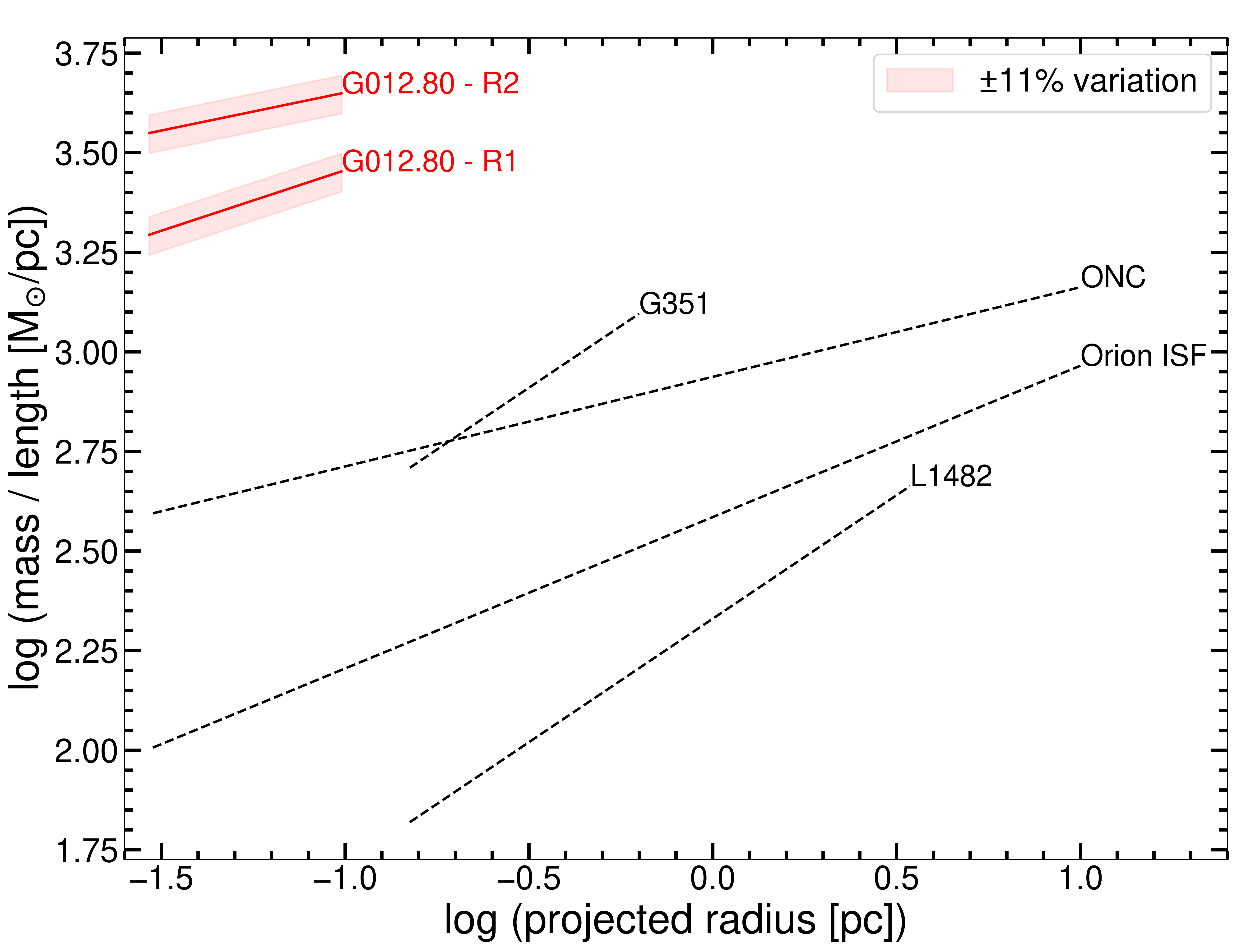}
    \vspace{-0.5cm}
    \caption{Line-mass profile of the R1 and R2 filaments (red lines). The red shaded area represents the uncertainty in both estimates, derived from the relative abundance uncertainty. This percentage therefore provides a lower limit to the error of 11$\%$ in both distributions. We include profiles of the Orion ISF \citep{stutz2016}, the ONC \citep{stutz2018}, the California L1482  region \citep{gutierrez2021}, and the G351.77 protocluster \citep{reyes2024} with black lines. The main filaments of G012 reveal denser line-mass distributions than those found in other star forming regions.}
    \label{fig:enclosedmass}
\end{figure}

Characterization of the mass distribution, and hence of the gravitational potential, is a requirement for interpreting the kinematics in these systems since the gravitational field of the gas serves as governor of the dynamics of a system \citep[e.g.,][]{stutz2018,gonzalez2019,gutierrez2021,reyes2024}. We follow the formalism outlined in \citet{stutz2016} to estimate these line-mass profiles in the main filaments of the G012 protocluster. We first align the filament mass maps with respect to their ridgelines (see Appendix \ref{sec:ap_datalign}). We use a box region to capture the densest parts of both filaments (see Fig.\ref{fig:SPITZER_N2H+filaments}), and construct \nthp cumulative mass distributions for both regions. As expected for elongated and dense structures, we find that the cumulative mass distribution profiles in both R1 and R2 are nearly linear (see Fig.\ref{fig:R1cumulativemass}), indicating that the mass distribution along the filament is approximately uniform. This allows us to derive an averaged line-mass profile as a function of projected radius from the ridgeline over the entire filament. While both distributions share this linear behavior, they differ in normalization: R2 is systematically shifted toward higher masses compared to R1. Following the formalism in \citet{stutz2016}, we find that the line-mass profiles are well-described by power laws of the form:

\begin{equation}
    \lambda(\omega) = \zeta \left( \frac{\omega}{\mathrm{pc}} \right)^{\gamma}
    \label{eq:linemassprofile}
\end{equation}
where $\omega$ is the projected radius in the POS (or impact paramenter from the ridgeline), $\gamma$ corresponds to the index in the enclosed mass over length (M/L) vs. projected radius diagram (see Fig.~\ref{fig:enclosedmass}), and $\zeta$ is the M/L normalization constant. For R1 and R2 we obtain:

\begin{equation}
\begin{split}
 \lambda_{\mathrm{R1}}(\omega) &= 5660 \hspace{0.5mm} \frac{\mathrm{M}_{\odot}}{\mathrm{pc}}\hspace{0.5mm} (\omega/\mathrm{pc})^{0.30}; \\ 
 \lambda_{\mathrm{R2}}(\omega) &= 6943 \hspace{0.5mm} \frac{\mathrm{M}_{\odot}}{\mathrm{pc}} \hspace{0.5mm} (\omega/\mathrm{pc})^{0.20}.
\end{split}
\end{equation}
In Fig.~\ref{fig:enclosedmass}, we display the M/L profiles of the R1 and R2 filaments (red and blue lines respectively), along with other star forming regions (black dashed lines) described below and in Table~\ref{tab:metrics}.

We follow \citet{stutz2016, stutz2018}; and \citet{gutierrez2021} to estimate the density, gravitational potential, and acceleration assuming cylindrical 3D geometry. All quantities are estimated in the POS since we do not have access to inclination information. We refer to these as ``apparent'' profiles following \citet{gutierrez2021}. We estimate the apparent volume density as:

\begin{equation}
    \rho(\mathrm{r}) = \beta \left( \frac{\mathrm{r}}{\mathrm{pc}} \right)^{\gamma - 2},
    \label{eq:volumedensity}
\end{equation}
where for the R1 and R2 filaments, we obtain $\beta_{\mathrm{R1}}$\,=\,211.3\,M$_{\odot}$\,pc$^{-3}$ and $\beta_{\mathrm{R2}}$\,=\,181.3\,M$_{\odot}$\,pc$^{-3}$. We estimate the gravitational potential as:

\begin{equation}
    \phi(\mathrm{r}) = \psi \left( \frac{\mathrm{r}}{\mathrm{pc}} \right)^{\gamma},
    \label{eq:gravitationalpot}
\end{equation}
where, considering the R1 and R2 parameters described above, we estimate $\psi_{\mathrm{R1}}$\,=\,123.1\,km$^2$\,s$^{-2}$ and $\psi_{\mathrm{R2}}$\,=\,269.3\,km$^2$\,s$^{-2}$. Finally, we estimate the gravitational acceleration as follows:

\begin{equation}
    g(\mathrm{r}) = -\xi \left( \frac{\mathrm{r}}{\mathrm{pc}} \right)^{\gamma - 1}.
    \label{eq:gravitationalacc}
\end{equation}
For R1 and R2 we obtain $\xi_{\mathrm{R1}}$\,=\,37.4\,km$^2$\,s$^{-2}$\,pc$^{-1}$ and $\xi_{\mathrm{R2}}$\,=\,51.3\,km$^2$\,s$^{-2}$\,pc$^{-1}$, respectively. 

Relative to other star forming regions (see Fig.~\ref{fig:enclosedmass}), the filaments in G012 exhibit significantly higher line-mass profiles, up to three times more dense than extended regions such as the G351.77 protocluster \citep{reyes2024}. This trend remains even when considering more extended regions along the filaments, as detailed in Appendix\,\ref{sec:ap_filfinder}. The corresponding line-mass profile parameters, including the filament extensions, are listed in Table~\ref{tab:metrics}.

\subsection{Star formation rate and efficiency}\label{sec:mass}

The star formation rate (SFR) and efficiency (SFE) are critical parameters to understand the current star formation activity, offering insight into the efficiency in which gas is converted into stars. To estimate the SFR (Eq.~\ref{eq:SFR}), we assume lifetimes based on observational studies of pre- and protostellar cores. Specifically, for the prestellar cores, their lifetimes vary significantly with environment and initial conditions, ranging from 0.5 to 2\,Myr in low-mass regions \citep[e.g.,][]{jessop2000,merin2008,evans2009,konyves2015} to as short as 0.05 - 0.24\,Myr in high-mass cores, where collapse is driven by fast converging flows \citep{csengeri2011a,csengeri2011b,motte2018,maxime2024}. As cores evolve into protostellar phases, accretion becomes dominant, and lifetimes decrease with increasing mass from $\sim$0.5\,Myr for low-mass protostars \citep{tom2022} to 0.1 - 0.3\,Myr for high-mass cores \citep{duarte2013,maxime2024}. Gravitational contraction, temperature increase, and accretion rates are strongly influenced by the core's initial mass and local conditions \citep{bontemps2010,dunham2014}.

For the SFR and SFE estimates, we use the formalism outlined in \citet{tom2022} and consider the G012 core masses provided in \citet{motte2024}. In R1 and R2, \citet{armante2024} and \citet{motte2024} identified a total of 16 cores. Fourteen of these are located in R2, and only two are in R1. These cores are classified as either being prestellar or protostellar, with masses ranging from 0.3 to 3\,M$_{\odot}$. For our analysis, we adopt a prestellar timescale of 1.2\,Myr, the expected value for prestellar cores in the mass range of 0.003 - 10\,M$_{\odot}$ \citep{konyves2015}. We also used protostellar lifetime of 0.5\,Myr, assuming a low-mass protostellar regime around 0.5\,M$_{\odot}$ \citep{tom2022}. Both cores in R1 are classified as prestellar, while the R2 filament exhibits five cores in this phase. In addition, the R2 filament contains nine protostellar cores. Considering the pre- and protostellar lifetimes described above, we estimate the SFR as follows:

\begin{equation}
    \mathrm{SFR}(\mathrm{M}_{\odot} \mathrm{Myr}^{-1}) = \frac{\mathrm{M}_{\mathrm{cores}}}{\mathrm{t}_{\mathrm{cores}}}.
    \label{eq:SFR}
\end{equation}
Here, $\mathrm{M}_{\mathrm{cores}}$ is the mass of the prestellar/protostellar core population and $\mathrm{t}_{\mathrm{cores}}$ is the approximate lifetime of the prestellar/protostellar cores in Myr. For R1 and R2, we obtain:

\begin{equation*}
\begin{split}
	&\mathrm{SFR}_{\mathrm{R1 - prestellar}} = (4.24 \pm 0.29) \hspace{1mm} \mathrm{ M}_{\odot}\mathrm{Myr}^{-1},\\
	&\mathrm{SFR}_{\mathrm{R2 - prestellar}} = (12.1 \pm 1.90) \hspace{1mm} \mathrm{M}_{\odot}\mathrm{Myr}^{-1},\\
	&\mathrm{SFR}_{\mathrm{R2 - protostellar}} = (43.2 \pm 6.67) \hspace{1mm} \mathrm{M}_{\odot}\mathrm{Myr}^{-1},\\
	&\mathrm{SFR}_{\mathrm{R2 - total}} = \mathrm{SFR}_\mathrm{prestellar} + \mathrm{SFR}_\mathrm{protostellar} = (55.3 \pm 6.93) \hspace{1mm} \mathrm{M}_{\odot}\mathrm{Myr}^{-1}.\\
\end{split}
\label{eq:SFR_val}
\end{equation*}
Errors were estimated solely from the mass uncertainties reported in the \citep{motte2024} catalog, without including possible uncertainties in the adopted timescales. For instance, adopting a lifetime of 0.3\,Myr for the intermediate to high-mass protostellar regime would increase the SFR to (71.9~$\pm$~11.2)~M$_{\odot}$~Myr$^{-1}$. 

In addition, we estimate the SFE as:

\begin{equation}
    \mathrm{SFE} = \frac{\mathrm{M}_{\mathrm{cores}}}{\mathrm{M}_{\mathrm{cloud}}} + \mathrm{M}_{\mathrm{cores}}.
    \label{eq:SFE}
\end{equation}
In this case $\mathrm{M}_\mathrm{{cores}}$ represents the same core mass used for the SFR estimate, and $\mathrm{M}_{\mathrm{cloud}}$ corresponds to the R1 or R2 filament mass (see Table~\ref{tab:fil}). In R1 and R2, we obtain:

\begin{equation*}
\begin{split}
	&\text{SFE}_{\text{R1 - prestellar}} = (1.6 \pm 0.1) \times 10^{-3},\\
	&\text{SFE}_{\text{R2 - prestellar}} = (3.3 \pm 0.4) \times 10^{-3},\\
	&\text{SFE}_{\text{R2 - protostellar}} = (4.8 \pm 0.4) \times 10^{-3},\\
	&\text{SFE}_{\text{R2 - total}} = \text{SFE}_\text{prestellar} + \text{SFE}_\text{protostellar} = (8.1 \pm 0.5) \times 10^{-3}.\\
\end{split}
\label{eq:SFE_val}
\end{equation*}
In this case, we adopted the same assumptions for the core mass uncertainties as for the SFR. For the cloud (filament) mass, we assumed an 11$\%$ error, corresponding to the uncertainty associated with the relative abundance values in the region. Our estimates of SFR and SFE should be regarded as lower limits, since they are based exclusively on the population of pre- and protostellar cores traced by some specific tracers. Unlike nearby star-forming regions, more distant regions such as G012 require more complete core censuses to provide accurate determinations. Overall, the SFR and SFE in R1 are roughly an order of magnitude lower than in R2, a trend that even persists when considering more extended filamentary regions (see Appendix \ref{sec:ap_filfinder}).

\section{A filament rotation toy model applied to R1}\label{sec:rotationR1}

In the \nthp PV diagram, the R1 filament shows a helical pattern, a feature also observed in the California L1482-S cloud and attributed to filamentary rotation \citep{gutierrez2021,hsieh2021}. In Fig.~\ref{fig:pv_sub} (bottom-right panel), we highlight this feature at the top of R1 which is also associated with a clear average-velocity gradient of $\sim$10.4\,km\,s$^{-1}$\,pc$^{-1}$ (see Fig.~\ref{fig:gradients}, left panel). 

To determine if rotation or gravity is the dominant factor in the R1 filament, we compare the velocity gradient to the gravitational acceleration implied by the filament mass distribution. We apply the formalism from \citet{gutierrez2021} to estimate the centrifugal force associated with rotation of the R1 filament. We then calculate the ratio between the centrifugal and gravitational forces as follows:

\begin{center}
\begin{equation}
	\frac{\mathrm{F_c}}{\mathrm{F_g}} = \frac{\mathrm{VG }^2 \mathrm{pc}^{-1}}{\xi \cos^3(\theta)} \left(\frac{\mathrm{r}}{\mathrm{pc}} \right)^{-\gamma}; \hspace{3mm} \mathrm{VG} = \text{10.4 } \frac{\mathrm{km}}{\mathrm{s pc}}, \hspace{1mm} \xi = \text{37.4 } \frac{\mathrm{km}^2}{\mathrm{s}^2 \mathrm{ pc}}.
\label{eq:FORCES}
\end{equation}
\end{center}

Here, $\mathrm{F_c}$ and $\mathrm{F_g}$ represent the centrifugal and gravitational forces, respectively. VG is the average-velocity gradient associated with the R1 filament rotation, $\xi$ is the constant associated with the gravitational acceleration distribution (see Table~\ref{tab:metrics}), and $\theta$ represents the (unknown) inclination angle of the filament relative to the POS.

\begin{figure}[h!]
    \includegraphics[trim=0cm 14cm 0cm 0cm, clip,width=1\linewidth]{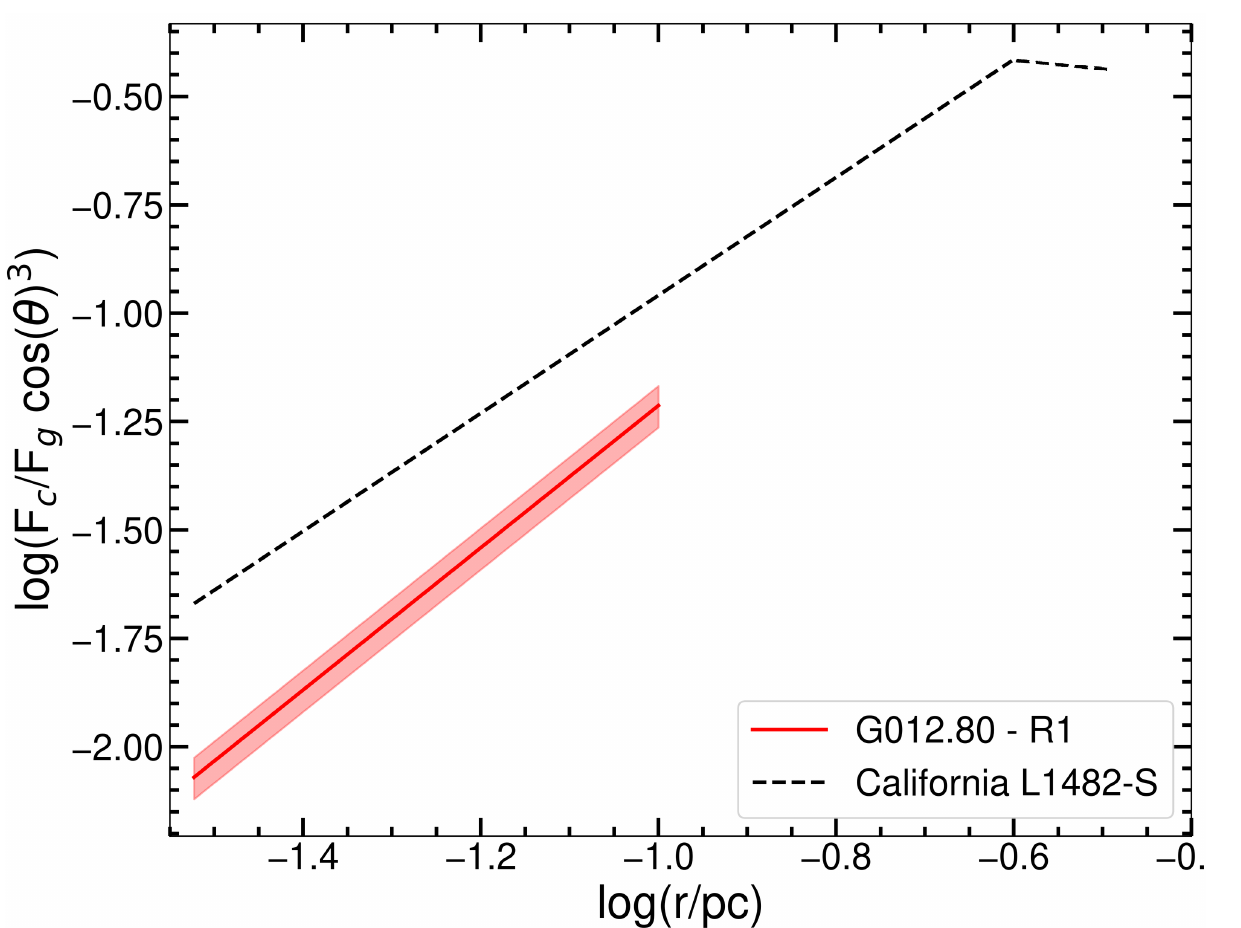}
    \vspace{-0.6cm}
    \caption{Ratio between the centrifugal ($F_c$) and gravitational ($F_g$) forces in the R1 filament (red line), where $\theta$ represents the inclination angle of the filament relative to the POS. We assume cos$(\theta)$\,=\,1, corresponding to a filament aligned with the POS. The red shaded area represents the lower limit error of 11$\%$, derived from the relative abundance uncertainty. For \nthp, the gravitational force dominates over rotational force, with a distribution similar to the internal rotation observed in the California L1482-S filament \citep{gutierrez2021}.}
\label{fig:R1forcesratio}
\end{figure}

In Fig.~\ref{fig:R1forcesratio}, we show the resulting profile for the ratio of forces in the R1 filament (red line), compared to the L1482-S profile (black line) in the California cloud \citep{gutierrez2021}. For this comparison, we assume $\cos(\theta)$=1, i.e., the filament is not inclined with respect to the POS. For reference, an inclination of $45\degree$ moves the R1 curve upward in this diagram by a factor of $\sim 0.45$, closer to the L1482-S curve. In other words, a non-zero POS inclination would increase the role of rotation in the filament. Assuming cos$(\theta)$\,=\,1, the R1 profile indicates that the filament is predominantly influenced by gravity rather than rotation, similar to the internal rotation pattern observed in the California L1482-S cloud. The high line-mass profile seen in the R1 filament, combined with the gravitational dominance in this profile, suggests that the filament could be primarily dominated by gravity, even in the presence of rotation. Thus its most likely fate is to collapse into a state similar to R2. 

Interestingly, despite the differences in global properties such as total mass and environment between R1 and the California L1482-S filament, both structures exhibit remarkably similar kinematic signatures, including helical patterns and comparable ratios between centrifugal and gravitational forces. This suggests that such  rotation-like features arise in structures filamentary covering a very large range in line-mass values. Hence, these signatures could reflect more general kinematic processes inherent to filament evolution, potentially making them a common feature in both low- and high-mass star-forming environments.

However, the features of potential rotation do not generically appear in the same line tracer always, although to date there are only two examples that we know of (and one is presented here, in \nthp). In contrast to California L1482-S, we do not observe rotation in the \ceto gas in the R1 filament. \nthp and \ceto trace different physical conditions within molecular clouds, reflecting variations in density, temperature, and chemistry that evolve during star formation \citep[e.g.,][]{tafalla2004,tanaka2013}. \ceto, a stable CO isotopologue, remains in the gas phase in relatively warm, low-density environments and thus traces material prior to significant CO depletion \citep[e.g.,][]{friesen2010,punanova2016,sabatini2022}. As clouds contract and cool, CO freezes onto dust grains, creating conditions that favor the survival of \nthp \citep{caselli2002b}. This chemical progression marks the transition from quiescent to dense, star-forming gas. Hence, rotation detected in different tracers provides clues to the evolutionary stage of a region. While L1482-S \citep{gutierrez2021} and R1 share similar kinematics and low core formation rates, their contrasting line-mass profiles suggest that R1 is in a slightly more advanced evolutionary phase.

\section{Discussion}\label{sec:dis}

Despite the morphological similarities between the two main G012 filaments, our analysis reveals differences in their gas kinematics, mass distributions, and core population. In the following discussion, we explore the different star forming scenarios that may be associated with the R1 and R2 filaments, emphasizing their potentially different evolutionary paths.

\begin{figure*}[h!]
    \centering
    \includegraphics[width=1\linewidth]{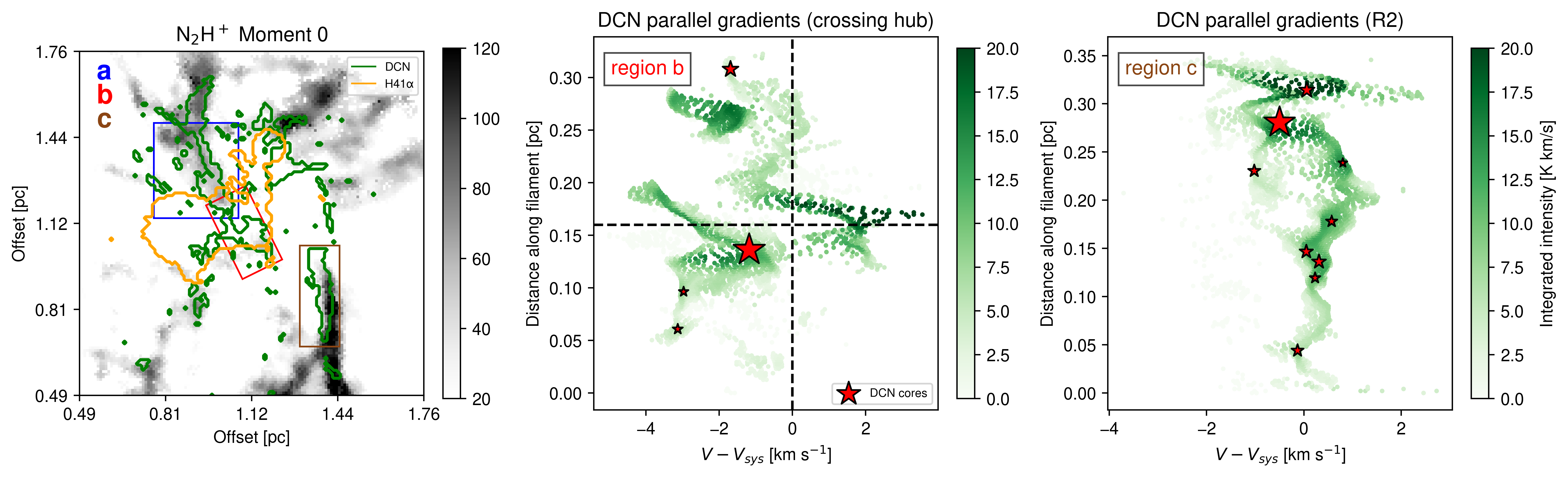}
    \vspace{-0.6cm}
    \caption{DCN PV diagrams along the R2 filament and toward the central \hh region. The left panel shows a zoom-in of the \nthp moment 0 map, with DCN (green) and \hfoa (yellow) contours overlaid. The middle and right panels present DCN PV diagrams extracted along the red and brown boxes shown in the left panel, respectively. The red box highlights the filament extending toward the central \hh bubble in the protocluster, while the brown box traces the R2 filament, including the northern region lacking \nthp emission. Red star symbols indicate the cores detected in DCN \citep{DCN}. In the middle panel, black dashed lines mark the center of the \hh bubble (horizontal line) and the systemic velocity estimated from masers \citep{immer2014}, corresponding to the approximate velocity of the \hh bubble (37\,$\kms$; vertical line).}
\label{fig:dcn_pv}
\end{figure*}

\subsection{\nthp destruction} \label{sec:nthpdestruction}

In Sec.~\ref{sec:analisis} we compared \nthp emission with complementary tracers related to ionization, outflows, and dense gas in the G012 protocluster. Specifically, we highlight a spatial anti-correlation between \nthp, \ceto emission, and the \hfoa recombination line. Previously, \citet{tobin2013} estimated the abundances of \nthp in protostellar systems. In one of these systems (L1157), \ceto traces regions where \nthp is absent. Specifically, they found that as \ceto peaks decrease, \nthp increases, indicating a clear anticorrelation between both tracers. This anticorrelation is also detected by \citet{tafalla2021}, where they also show that \nthp is detected at high column densities (N$(\text{H}_2)\,\geq\,10^{22}\,$cm$^{-2}$; see their Fig.~7). In addition, \citet{yu2022} analyzed a \hh bubble (S156) associated with the G305 star forming complex, finding that \nthp increases far to the edges of S156. This implies that \nthp could be destroyed by the hot and expanding gas in the bubble. As well as increased CO near \hh regions, free electrons traced by recombination lines (e.g., H41$\alpha$), ionized metal emission lines, and continuum emission provide additional methods to map these zones where \nthp is destroyed \citep[e.g.,][]{yu2018,continuum,armante2024}.

\citet{roberto2024} analyzed the 3\,mm continuum emission and \hfoa recombination line data for the full ALMA-IMF sample, finding that an increase in emission is mostly seen in evolved regions where OB associations exist. All of these studies, along with the effect of temperature on the critical density of the \nthp molecule (ranging from 6.1\,$\times$\,10$^4$\,cm$^{-3}$ at 10\,K to 2.0\,$\times$\,10$^4$\,cm$^{-3}$ at 100\,K), suggest a close relationship between the decrease in \nthp and the evolutionary stages of the protocluster. The critical density indicates the conditions under which \nthp becomes an effective tracer, thus linking its observed decrease to changes in the protocluster condition. In G012, the 1.3\,mm continuum emission is concentrated at the center of the region and to the north of the R2 filament (see Fig.\,\ref{fig:SPITZER_N2H+filaments}, right panel). Moreover, the 3\,mm continuum exhibits high-flux intensity distributions appearing as bubbles in the protocluster center \citep{continuum}, similar to the \hfoa distribution (bottom-left panel of Fig.~\ref{fig:TRACERS}). In addition, we note a decrease in the \nthp relative abundance distribution in regions near to \hfoa bubbles and ionizing emision from hot cores, highlighting the impact of feedback mechanisis in the distribution and abundance of the dense gas traced by \nthp. Given the high concentrations of \hfoa and \ceto in the center of G012, it is likely that the central \hh regions are heating the gas in the protocluster and destroying the \nthp molecule through interactions with free electrons \citep{vigren2012} and with the CO molecule \citep{bergin1997,caselli2002b,caselli2012}.

\subsection{DCN accretion evidence in the R2 filament and towards the central OB star cluster}

As \nthp is significantly affected by stellar feedback, particulary by the increase in temperature associated with the central massive OB star cluster, we further investigate the kinematics of the gas in these regions using DCN emission, which is more resilient to temperature enhancements. Although DCN does not trace filamentary structures as extended as those revealed by \nthp, it highlights hotter and denser gas components, and reveals filamentary features within the R2 filament and in structures located close to the central OB cluster and respective \hh bubble.

First, we assess whether the R2 filament shows evidence of ongoing core accretion processes. Recently, \citet{pierre2024} derived a dust temperature map of the G012 protocluster, finding values of $\sim$~35\,K at the top of R2. They also reported a spatial shift between the \hdos and \nthp column density peaks, which they interpret as the effect of a detected hot core \citep{armante2024} heating the gas and destroying \nthp at the top of the filament. DCN emission in this region reveals a clear ``V-shape'' structure, features seen in other filaments undergoing accretion \citep[e.g.,][]{liu2019,bonne2020,bonne2023,yiyo2024,nico2024}, coincident with the hot core location (see Fig.\,\ref{fig:dcn_pv}, right panel), along with an SiO elongated feature perpendicular to the filament related with an outflow component \citep[see Fig.\,3 in][and Fig. \ref{fig:TRACERS}]{armante2024}. Additional kinematic evidence is found along the filament. As shown in Fig.~\ref{fig:gradients} (right panel), the average-velocity diagram reveals a clumping of velocities around $\sim$\,1\,$\kms$ in the central and densest region of R2, spatially coincident with the integrated intensity peak. Such clumps in PV space are commonly interpreted as signatures of accretion-driven core formation \citep[e.g.,][]{hacar2013,kainulainen2016,ladjelate2020}, marking regions where gravity overcomes thermal support \citep[e.g.,][]{andre2010,palau2013}. Thus, the observed velocity clumping in addition to the small velocity spreads in the PV diagrams features along the R2 filament and the similar values in velocity between cores and dense gas further points to an active role of cores in shaping the filament kinematics.

To place R2 in context, we compare its star-forming properties with those of the ISF OMC-2 filament. From previous catalogs \citep{tom2012,tom2016,furlan2016,kainulainen2017,stutz2018,gonzalez2019} we derive a SFR of 95\,M$_{\odot}$\,Myr$^{-1}$ and a SFE of 0.008 for protostellar cores in OMC-2. In addition, previous studies estimated a line-mass profile of $\lambda(\omega) = 385\,M_{\odot}$\,pc$^{-1}$($\omega$/pc)$^{0.38}$ \citep{stutz2018}. Although R2 shows an even steeper line-mass profile, its SFR is lower than that in OMC-2 (see Fig.~\ref{fig:enclosedmass}), while both regions exhibit similar efficiencies. This comparison suggests that R2 is already in an evolved phase of filamentary accretion, possibly slightly earlier than OMC-2 but progressing toward a comparable state. This makes R2 an excellent candidate for testing filamentary accretion scenarios with upcoming JWST observations of YSOs in these filaments.

In addition to the accretion evidence identified in the R2 filament, in Fig.\,\ref{fig:dcn_pv} we show a set of coherent velocity structures traced by DCN emission extending across the southern portion of R1 (blue box in the left panel) and toward the central region of the protocluster where the OB cluster is located (red box in the left panel). These structures appear as elongated features in DCN (green contours in the left panel) and display organized velocity patters in parallel PV diagrams along to the central region (middle panel). In particular, the parallel PV diagrams reveal multiple velocity gradients that form characteristic V-shaped patterns. While such V-shaped structures are often interpreted as signatures of gas inflow, we note that similar PV morphologies can arise from different kinematic configurations (e.g., rotation or outflows), and therefore do not uniquely trace accretion. One of these structures is detected toward the southern extended portion of the R1 filament (blue box). Here, DCN traces a relatively shallow V-shape that coincides spatially with a similar, though weaker, velocity pattern seen in the \nthp emission. Unlike the more prominent structures discussed below, this feature does not appear to be directly associated with a compact core. This may indicate that the V-shaped structure traces gas motions along the filament that are not associated with a compact and localized gravitational potential (e.g., a prestellar or protostellar core, as observed in the R2 filament), and could be consistent with large-scale gas flows along the southern portion of R1, potentially including inflow.

More promitent V-shaped structures are observed closer to the central region of the cloud (red boxes). Specifically, in DCN we find two well-defined V-shaped patterns (see Fig.\,\ref{fig:dcn_pv}, middle panel), where one of these appears to converge toward the position of a massive core detected in \citet{DCN}. The coincidence between the vertex of the V-shaped and the location of the core suggests that this velocity gradient may trace gas flows feeding this object along the filamentary structure \citep{yiyo2024,nico2024}. A second V-shaped velocity structure is detected nearby, with it's vertex lying closer to the central region associated with the OB star cluster (represented by a horizontal black dashed line in the figure). The projected position of this feature is consistent with the location of one of the \hh ionized bubbles in G012. While the spatial proximity to the ionized region raises the possibility that this observed gradient could be influenced by stellar feedback, several observational characteristics suggest that it is more consistent with inflow motions rather than outflowing gas. In particular, the relatively low velocity variations (around 4\,$\kms$) are significantly smaller than those typically observed in outflows, and the emission is traced by DCN, which probes cold and dense gas not usually associated with high-velocity ejected material. Together, these properties favor an interpretation in terms of gas motions directed toward the central \hh\ region, possibly related to inflow, although contributions from feedback-induced perturbations or other complex velocity configurations cannot be entirely ruled out. We higlight that such kind of patterns that conect R1 with the central region are not found for the R2 filament in \nthp or DCN.

\subsection{Potential star-forming scenarios in G012}

\begin{figure*}[h!]
    \centering
    \includegraphics[trim=0.0cm 21cm 0cm 0cm, clip,width=0.9\linewidth]{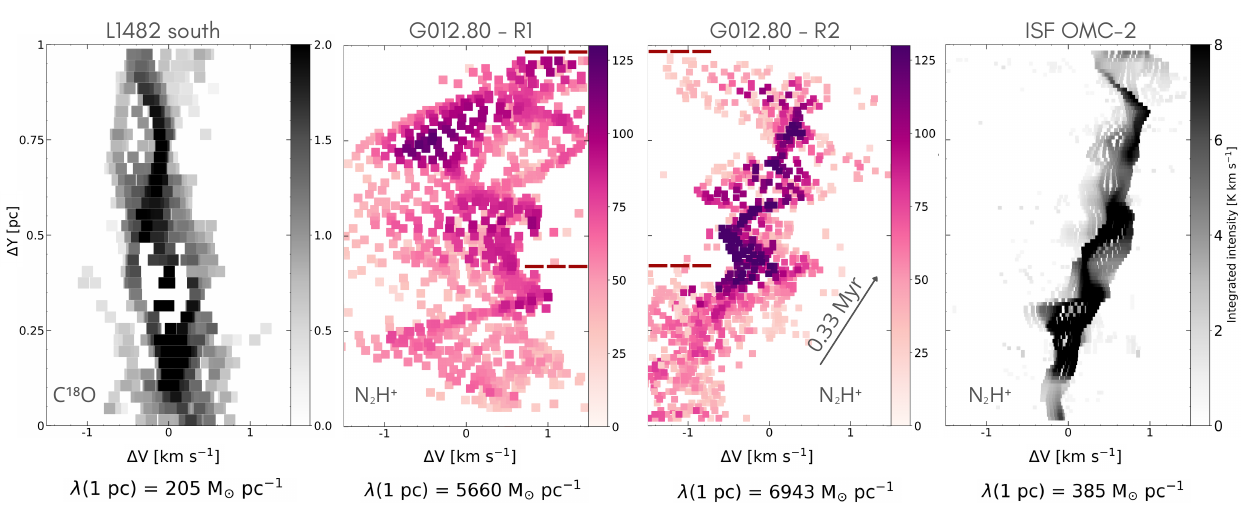}
    \vspace{-0.3cm}
    \caption{Comparative PV diagrams of California L1482-south \citep[left panel,][]{gutierrez2021}, OMC-2 \citep[right panel,][]{gonzalez2019}, R1 and R2 filaments (center panels). Similar to California, R1 presents a ``double helix'' feature, low presence of cores, and low temperatures \citep[< 25\,K, see][]{pierre2024}. Despite this, line-mass profile of both regions (estimated within the range of the red lines on the central panels) shows that R1 is more massive and dense than California L1482-south. Furthermore, R2 shows PV diagram features comparable to the ISF OMC-2 region. Specifically, R2 presents smooth undulations in velocities characterized with uniform velocities. R2 is also characterized by a high presence of cores, indicating that the gas velocities should be mostly influenced by the accretion of these sources.}
\label{fig:evolution}
\end{figure*}

The formation of stars occurs over diverse timescales, reflecting the intricate processes governing their evolution \citep{bergin2007,evans2009,dunham2012,dunham2014,motte2018}. These variations have direct implications for the G012 protocluster. In the R1 filament, which contains only two prestellar cores \citep{armante2024}, the low SFR of $\sim$\,4.24~M$_{\odot}$~Myr$^{-1}$ and an SFE of $\sim$\,0.002 suggest that star formation is proceeding slowly despite the high gas density. In contrast, the R2 filament contains both prestellar and protostellar cores and shows a significantly higher SFR of $\sim$\,55.3~M${\odot}$~Myr$^{-1}$ and an SFE of $\sim$\,0.008 (see Sec.~\ref{sec:mass}). Although these estimates are derived from subregions (R1 and R2) selected to isolate the kinematically coherent portions of the filaments, rather than their full spatial extent, the same trend is also observed when considering the larger scale structures (see Appendix \ref{sec:ap_filfinder}): the extended R1 filament still shows lower SFR and SFE values than the extended R2 filament.

The presence of multiple protostellar cores indicates that R2 has been more active in forming cores up to now and is a more efficient star-forming region. The lack of protostellar cores and very limited number of prestellar cores in R1, together with the kinematic differences previously reviewed, point to marked contrasts between the two filaments, suggesting that R1 is at an earlier evolutionary stage, prior to the filament concentration and active star formation already evident in R2. This implies that, while potentially less evolved than R2, the R1 filament may eventually reach a higher level of core formation as its dense gas reservoir becomes more active. However, the physical processes driving this activity remain uncertain. Similar contrasts in core populations have also been reported in young filaments, such as those in the G351 protocluster \citep{nico2024}. However, the extended structure of R1 suggests that its internal evolutionary state may not be uniform. In particular, the lower portion of the filament, located closer to the central region hosting the OB star cluster and its associated \hh bubbles, shows a higher concentration of dense cores and more complex kinmatic signatures than the upper part of the R1 filament. The DCN PV diagrams reveals several V-shaped velocity patterns in the extended area at the bottom of R1, some of which converge toward massive cores while others appear spatially connected to the central ionized region. These features may trace localized gas inflows along filamentary structures toward either individual cores or the cluster enviroment. One possible interpretation is that the expansion of the nearby \hh bubble is compressing the surronding molecular gas, potentially enhancing core formation in the lower portion of R1. Such feedback driven triggering has been suggested in other star-forming regions where expanding ionized bubbles interact with dense filaments \citep[e.g.,][]{doris2022,neupane2024}. In this context, the higher abundance of dense cores in this bottom and extend part of R1 may reflect a locally accelerated phase of star formation (compared to the upper part that exhibits potential rotation signatures) superposed on an otherwise less evolved filament.

Additional evidence supporting this evolutionary contrast comes from chemical differentiation in other regions where molecular tracers such as DCN and N$_2$D$^+$ highlight different physical conditions \citep[e.g.,][]{sakai2022,DCN}, suggesting that similar mechanisms could be at play in G012. Moreover, the estimated lifetimes of high-mass prestellar cores in \citet{maxime2024} are comparable to the velocity gradient timescale in R1 ($\sim$\,0.1\,Myr), suggesting that the observed kinematic conditions may set a characteristic timescale for core formation. This supports a causal view where gas kinematics drive core assembly and evolution, rather than simply resulting from it. However, this interpretation likely applies primarly to the filament regions that are not strongly affected by external feedback. In the lower part of the extended R1 structure (only the region traced by DCN), which lies closer to the central \hh region, the gas kinematics may also be influenced by the expansion of the ionized bubble, potentially modifying the local conditions for core formation (see above). The contrasting core populations and timescales in R1 and R2 thus highlight how local dynamical conditions regulate the star formation efficiency and rate. Comparisons with other regions reinforce this evolutionary interpretation. As shown in Fig.~\ref{fig:evolution}, the PV-diagram features and line-mass profiles of R1 and R2 resemble those of filaments at different stages: R1 shares characteristics with the early-stage, quiescent L1482-S region \citep{gutierrez2021}, while R2 resembles the more evolved Orion ISF \citep{gonzalez2019}. For this comparison to be meaningful, it's important to verify that the datasets probe similar physical and kinematics scales. Although the angular resolutions differ among the above mentioned datasets, the physical resolutions are comparable. The \nthp observations of \cite{gonzalez2019} and \cite{gutierrez2021} have spatial resolutions of 17-22" at distances of 420-500 pc, corresponding to ~7-11\,kau, respectively. Our observations have a beam of ~2.3" at 2.4 kpc, corresponding to ~6\,kau. Therefore, all datasets probe similar physical scales along the filaments. The spectral resolution in our data (0.23 $\kms$) is moderately larger than in those studies (0.12-0.16 $\kms$), which may smooth very fine velocity substructure but doesn't affect the identification of global velocity gradients or velocity spread features discussed here.

The coexistence of filaments in distinct evolutionary phases within the same protocluster has been observed in other regions \citep[e.g.,][]{tafalla2004,andre2010,schneider2012,hacar2013,peretto2014,zhou2022,nico2024}. For instance, in the Taurus cloud, \citet{hacar2013} identified coherent filaments exhibiting both quiescent and active regions. Similar findings were reported in SDC13 \citep{peretto2014} and the Rosette cloud \citep{schneider2012}, where filaments with different densities, temperatures, and fragmentation levels coexist. Several factors can explain this diversity, including initial density variations, differential gas accretion, and local feedback from protostellar outflows and \hh. Denser filaments tend to collapse earlier, while diffuse structures may remain stable or be affected by nearby activity. These mechanisms can lead to asynchronous star formation, where young stellar objects in the same cluster span a wide range of ages \citep{peretto2014,nony2021}. Recognizing the presence of filaments at different stages within a single region like G012 is key to understanding the temporal and spatial complexity of star formation in protoclusters. An intriguing open question is whether the apparent rotation of R1 reflects a genuine large scale rotational motion with potential inherited angular momentum from the parent molecular cloud or arises from other kinematic configurations. Several alternative scenarios may produce apparent transverse velocity gradients:

\begin{itemize}
	\item A coherent rotation of the filament would naturally produce a smooth and approximately linear velocity gradient perpendicular to the filament ridgeline. This behavior is indeed observed in the R1 filament, where the ridgeline-averaged velocity vs. radius diagram reveals a well defined and continuous gradient (Fig. \ref{fig:gradients}, left panel).

	\item The observed gradient could result from unresolved dense substructures, such as fibers, forming cores, or accretion streamers. Simulations and observations suggest that young filaments may retain rotational signatures from their fragmentation process, with cores often exhibiting rotation axes preferentially aligned relative to the filament's orientation \citep{kong2019,hsieh2021,lee2025}. Alternatively, sub-filamentary networks such as fibers and streamers, common in turbulent clouds, can serve as conduits for organized flows and potentially impart angular momentum to host filaments \citep{hacar2017,maud2017,hacar2018,pineda2023,olguin2025}. In this case, one would expect a more fragmented velocity structure, with discontinuities or localized deviations along the filament. While such substructures may be present below the current angular resolution, the smoothness of the averaged gradient in the top of R1 suggests that they do not dominate the large scale kinematics.
	
	\item Organized inflows of gas along sub-filamentary networks can also imprint velocity gradients. These flows typically produce asymmetric or localized velocity patterns such as V-shaped structures in parallel PV diagrams. In R1 we analyzed parallel gradients but we do not find V-shaped structures in the region that exibits the proposed rotation feature. We only find a weaker V-shaped pattern that is detected in the bottom, extended portion of R1 traced by DCN and \nthp, and near the central \hh region. This may suggest localized inflows toward the central \hh bubble environment rather than along the entire filament (Fig. \ref{fig:dcn_pv}, middle panel).
	
	\item External feedback, such as the expansion of nearby \hh regions, could strongly perturb filament kinematics by compressing gas and enhancing velocity dispersions \citep{zhou2022,doris2022,neupane2024,zhao2025}. Although such effects may contribute locally, the absence of a comparable transverse gradient in the nearby R2 filament, which lies within the same large scale environment, suggests that feedback alone cannot explain the kinematic behavior observed in R1.
\end{itemize}

Among the possible explanations, a coherent rotation of the filament provides the most straightforward interpretation of the observed transverse velocity gradient. While localized inflows or feedback related perturbations may contribute to the kinematics in specific regions, particularly in the bottom and extended portion of R1 near to the \hh region, the overall smoothness and continuity of the transverse gradient along the filament are most consistent with a large scale rotational motion. Disentangling these other alternative scenarios will require higher resolution observations capable of resolving sub-pc structures such as fibers, forming cores, and accretion streamers, as well as detailed comparisons with simulations of feedback influenced filaments.

\section{Conclusions}\label{sec:conclusion}

G012 is one of the most evolved and massive regions in the ALMA-IMF protocluster sample. Its dense gas emission, traced by the \nthp molecule, reveals a morphology mainly composed of two large filaments, and additional irregular but similarly dense structures. The center of G012 presents an absence of \nthp, in line with the existence of large-scale \hh regions, which destroys the dense gas forming a disrupted structure in the protocluster center. In this research we focus on the kinematical analysis of the \nthp emission of the G012 protocluster, including line-mass distribution analysis, and the kinematic characterization of its two main filamentary structures, which we name R1 and R2. Our main conclusions are summarized as follows:

\begin{enumerate}
\item We find multiple velocity components in G012. Around 55$\%$ of the \nthp spectra are well fitted by two velocity components, while the remainder are well described by a single velocity component. We define the two main velocity structures in the protocluster as the first and second velocity components, FVC and SVC respectively.

    \item The integrated intensity map reveals two dominant filamentary structures: the R1 and R2 filaments. Despite their similarity in morphology, we find obvious differences in their kinematics.
    
    \item R1 is characterized by a velocity gradient across its spine in the moment~1 map. We review the PV diagram features which showed a helical structure at the top of the filament, potentially associated with filament rotation. We find an average velocity gradient of $\sim$~10.4~$\kms$~pc$^{-1}$ (or $\sim$~0.1~Myr timescale) in this feature.
    
    \item R2 is characterized by smooth, compact velocity undulations (<~2~$\kms$) and the absence of a large scale velocity gradient.
    
    \item The mean line-mass profiles of both filaments are well described by $\lambda$($\omega$)~=~5660~M$_{\odot}$~pc$^{-1}$($\omega$/pc)$^{0.30}$ and $\lambda$($\omega$)~=~6943~M$_{\odot}$~pc$^{-1}$($\omega$/pc)$^{0.20}$ for R1 and R2, respectively. When we compare these distributions with those from other well-studied regions, we find that R1 and R2 stand out as having significantly higher line-mass profiles, consistent with having higher densities than the structures mentioned above. Based on these line-mass profiles, we calculate the gravitational field, force, and potential for both R1 and R2, assuming cylindrical symmetry.

    \item Using simple toy models, in R1 we estimate the ratio of centrifugal to gravitational forces (the latter based on the line-mass profile). We find that rotation in the R1 filament is subdominant compared to gravity, but as the radius increases the role of rotation increases, very similar to the behavior reported in the California L1482 filament \citep{gutierrez2021}
    
    \item While the kinematic features appear similar between L1482 and R1, the line-mass profile normalization at 1\,pc of R1 is almost 30 times higher than that of L1482. Hence, the similarities in the kinematics and the force-ratio profiles may indicate common angular momentum evolution of filaments across large ranges in mass and line-mass scales.  
    
    \item In terms of the estimated star formation activity, R1 lacks protostellar cores, and the few cores present are in the prestellar phase, implying an SFR~$\sim$~4.24~M$_{\odot}$~Myr$^{-1}$. In contrast, R2 hosts a larger population of massive cores compared to R1 and contains an SiO outflow oriented perpendicular to the filament. Specifically, R2 forms protostellar cores at a rate of SFR~$\sim$~55.3~M$_{\odot}$~Myr$^{-1}$ with a similar efficiency as the Orion ISF.
   
	\item We identify filamentary structures traced by DCN exhibiting signatures of potential accretion toward the center of the protocluster, where the OB stellar cluster is located and N$_2$H$^+$ is largely destroyed, suggesting that these structures may represent the remnants of a former hub-filament system.
	
    \item Using N$_2$H$^+$\,(1-0) and DCN\,(3-2) spectral lines, we estimate updated velocities for 48 previously identified cores. In general, the cores appear kinematically coupled to the dense gas traced by \nthp and DCN from which they are forming.
    
    \item Considering the R1 versus R2 differences in gas kinematics, core incidence, and estimated SFRs, we propose that R1 is still rotating and younger than the R2 filament, which has collapsed to a more efficient mode of star formation, likely by shedding its angular momentum.
\end{enumerate}

The results presented in this study emphasize the role of dense filamentary structures in high-mass star formation, especially when found at different stages of evolution within the same protocluster. Detecting filaments at different evolutionary phases within a single protocluster allows us to characterize multiple stages of star formation, from the earliest accretion processes to later stages of core collapse but after feedback effects from \hh regions. By uncovering these structures, we gain valuable insights into the diverse mechanisms governing the gas kinematics, including gas accretion and filament rotation evolution. This ability to observe a ``snapshot'' of the different stages in one region presents a unique window into the ongoing processes shaping Milky Way protoclusters.

\begin{acknowledgements}
ADS/JAO.ALMA$\#$2017.1.01355.L. ALMA is a partnership of ESO (representing its member states), NSF (USA) and NINS (Japan), together with NRC (Canada), MOST and ASIAA (Taiwan), and KASI (Republic of Korea), in cooperation with the Republic of Chile. The Joint ALMA Observatory is operated by ESO, AUI/NRAO and NAOJ. The project leading to this publication has received support from ORP, which is funded
by the European Union’s Horizon 2020 research and innovation program under grant agreement No. 101004719 [ORP]. A.S., J.S., N.S.G., R.A.G., N.C.T., and G.B.M. gratefully acknowledge support by the Fondecyt Regular (project code 1220610), and ANID BASAL project FB210003. R.G.M. and J.S. acknowledge support from UNAM-DGAPA-PAPIIT project IN105225. R.A.G. gratefully acknowledges support from ANID Beca Doctorado Nacional 21200897. F.L. and F.M. acknowledge funding from the European Research Council (ERC) via the ERC Synergy Grant ECOGAL (grant 855130) and from the French Agence Nationale de la Recherche (ANR) through the project COSMHIC (ANR-20-CE31-0009). A.K. gratefully acknowledges support from Fondecyt postdoctoral grant. L.B. gratefully acknowledges support by the ANID BASAL project FB210003. N.S.G. gratefully acknowledges support from ANID Beca Doctorado Nacional 21250244. G.B. acknowledges support from the PID2023-146675NB-I00 (MCI-AEI-FEDER, UE) program.

\end{acknowledgements}

\subsubsection*{Affiliations}
\begin{enumerate}
	\small{
     \item Departamento de Astronomía, Universidad de Concepción, Casilla
160-C, 4030000 Concepción, Chile  
       		
      \item Instituto de Radioastronomía y Astrofísica, Universidad Nacional
Autónoma de México, Morelia, Michoacán 58089, México \email{j.salinas@irya.unam.mx}
		
		\item Franco-Chilean Laboratory for Astronomy, IRL 3386, CNRS and Universidad de Chile, Santiago, Chile

		\item Scottish Universities Physics Alliance (SUPA), School of Physics and Astronomy, University of St. Andrews, North Haugh, St. Andrews KY16 9SS, UK

       \item Univ. Grenoble Alpes, CNRS, IPAG, 38000 Grenoble, France
       
    	\item INAF-Osservatorio Astrofisico di Arcetri, Largo E. Fermi 5, I-50125 Firenze, Italy
    	
    	\item Laboratoire d'Astrophysique de Bordeaux, Univ. Bordeaux, CNRS, B18N, all\'ee Geoffroy Saint-Hilaire, 33615 Pessac, France
    	
    	\item Department of Astronomy, University of Florida, PO Box 112055, Florida, USA

    	\item Departamento de Astronomía, Universidad de Chile, Las Condes, 7591245 Santiago, Chile

    	\item Department of Astronomy, School of Science, The University of Tokyo, 7-3-1 Hongo, Bunkyo, Tokyo 113-0033, Japan
    	
    \item Instituto de Astrofísica de Andalucía (CSIC), Glorieta de la Astronomía s/n, 18008, Granada Spain
     
     \item Departament de Física Quàntica i Astrofísica (FQA), Universitat de Barcelona, Martí i Franquès 1, E-08028 Barcelona, Catalonia, Spain
     
     \item Institut de Ciències del Cosmos (ICCUB), Universitat de Barcelona, Martí i Franquès 1, E-08028 Barcelona, Catalonia, Spain
     
     \item Institut d’Estudis Espacials de Catalunya (IEEC), Esteve Terradas 1, edifici RDIT, Parc Mediterrani de la Tecnologia (PMT) Campus del Baix Llobregat—UPC 08860 Castelldefels (Barcelona), Catalonia, Spain
     
     \item Departments of Astronomy and Chemistry, University of Virginia, Charlottesville, VA 22904, USA
     
     \item Instituto Argentino de Radioastronomía (CCT-La Plata, CONICET;
UNLP; CICPBA), C.C. No. 5, 1894, Villa Elisa, Buenos Aires, Argentina

	\item Laboratoire de Physique de l’École Normale Supérieure, ENS, Université PSL, CNRS, Sorbonne Université, Université de Paris, Paris, France

}

\end{enumerate}

\appendix

\section{N$_2$H$^{+}$ (1$-$0) line fitting: one and two velocity components} \label{sec:ap_linefitting}

In this appendix, we describe the data preparation steps prior to line fitting, including $\mathrm{S/N}$ analysis and input parameter selection. We also characterize the output modeling parameters and associated uncertainties, identifying the velocity structures used in the above analysis.

\subsection{Data preparation}\label{sec:data_prep}

Before fitting, we evaluate the robustness of the \nthp data cube by inspecting negative bowls and noise levels. The most negative values are concentrated toward the compact central region of G012. To identify reliable spectra, we construct an $\mathrm{S/N}$ map, where the noise is estimated pixel-by-pixel from the RMS in emission-free channels between 12-20\,km\,s$^{-1}$ and 49-61\,km\,s$^{-1}$. The $\mathrm{S/N}$ ratio is defined as the peak intensity per pixel divided by its RMS noise. We adopt an $\mathrm{S/N}$ threshold of 12, preserving most of the \nthp structure while excluding noisy or unresolved spectra unsuitable for fitting. Preliminary moment 0 and 1 maps are constructed using the full line for the integrated intensity and the isolated component for the mean velocity \citep[see][for details on isolated component identification]{yiyo2024}. The integrated intensity map (right panel in Fig.~\ref{fig:SPITZER_N2H+filaments}) reveals two main \nthp filaments and a disrupted central morphology in G012. The moment 1 map shows a velocity gradient perpendicular to the R1 filament ridgeline (Sec.~\ref{sec:gradients}) and broad velocity dispersion in the central region, likely associated with multiple velocity components.

\begin{figure}[h!]
    \centering
    \includegraphics[trim=5cm 0cm 0cm 0cm, clip,width=0.85\linewidth]{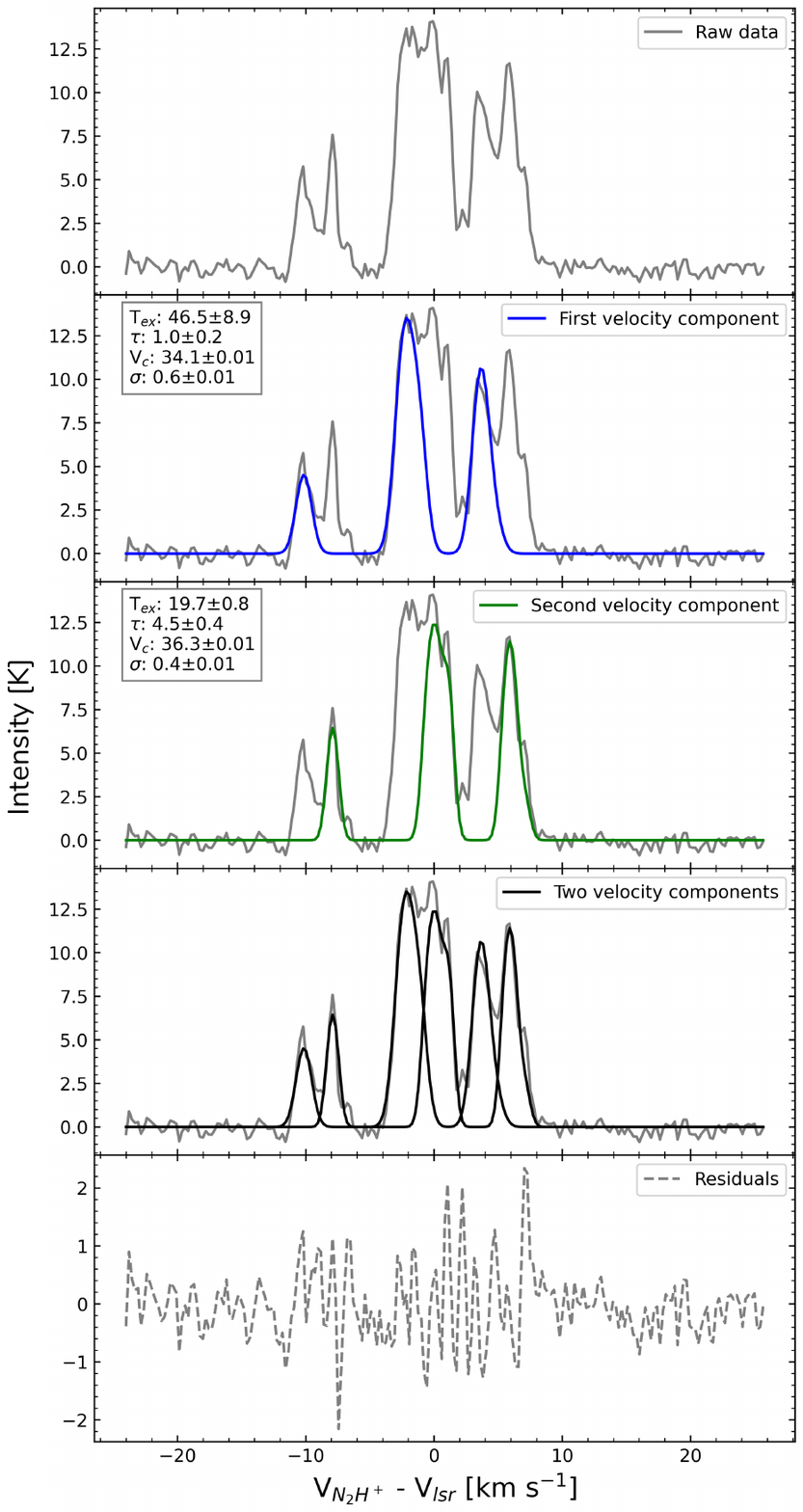}
    \vspace{-0.3cm}
    \caption{\nthp two velocity component fitting example: The panels, from top to bottom, display the raw data (grey curve), the first velocity component (blue curve) overlaid on the raw data, the second velocity component (green curve) overlaid on the raw data, the total model (black curve) fitted to the raw data, and the residuals of the model (grey dashed curve). The \texttt{PySpecKit} parameters for both the first and second velocity components are shown in each respective panel.}
    \label{fig:pyspeckit_ex}
\end{figure}

\subsection{Choosing input parameters}\label{sec:input_par}

\begin{table*}[h!]
\caption{Starting guesses for spectral fitting}
\vspace{-0.5cm}
\begin{center}
\begin{tabular}{ccccccccc} \hline \hline
Parameters & T$_{\mathrm{ext}}$(1) & $\tau$(1) &  $\mathrm{V_c}$(1) &  $\sigma$(1) & T$_{\mathrm{ext}}$(2) & $\tau$(2) &  $\mathrm{V_c}$(2) &  $\sigma$(2) \\ 
 & [K] &  &  [$\kms$] &  [$\kms$] & [K] &  &  [$\kms$] &  [$\kms$]\\ \hline  
\multicolumn{9}{c}{One velocity component} \\  \hline 
Guesses & 31.3 & 3.8 & 37 & 0.78 & --- & --- & --- & --- \\
Limits & (2.8, 150) & (0.001, 10) &  (25, 45) &  (0.23, 5) & --- & --- & --- & --- \\
Limited & (T,T) & (T,F) & (T,T) & (T,F) & --- & --- & --- & --- \\ \hline 
\multicolumn{9}{c}{Two velocity components} \\ \hline
Guesses & 15 & 1 & 33 & 1.4 & 20 & 3 & 37 & 0.47 \\
Limits & (2.8, 150) & (0.001, 10) & (25, 45) & (0.23, 5) & (2.8, 150) & (0.001, 10) & (25, 45) & (0.23, 5) \\
Limited & (T,T) & (T,F) & (T,T) & (T,F) & (T,T) & (T,F) & (T,T) & (T,F) \\ \hline
\end{tabular}
\end{center}   
\label{tab:guesses}
\end{table*}

To model the \nthp data, we used the \textit{specfit} fitting tool based on the built-in \textit{n2hp\_vtau} fitter. The \textit{specfit} task requires the following input values: \textit{fittype} specifies the model type and requires the number of velocity components and model parameters; \textit{guesses} are initial model parameter values (only applied to the first pixel fit); \textit{limits} are lower and upper parameter bounds used across all pixels via the Levenberg-Marquardt algorithm optimizing the $\chi^2$ function; \textit{limited} applies or disregards the \textit{limits} defined above; \textit{errmap} is the error map, calculated as the standard deviation in the noise channels (see above); \textit{signal\_cut} defines the signal lower limit for the modeling, where $\mathrm{S/N}$\,$>$\,\textit{signal\_cut}; and \textit{start\_from\_point} is the initial pixel for fitting.

We test models with one and two velocity components, using four main parameters: the excitation temperature (T$_{\mathrm{ex}}$), optical depth ($\tau$), velocity centroid (V$_{\mathrm{c}}$), and velocity dispersion ($\sigma$). For the initial $guesses$ and $limits$, we select values based on the preliminary moment maps described above. The \textit{start\_from\_point} is chosen as a pixel with high $\mathrm{S/N}$ and a well-defined spectrum. Testing various \textit{signal\_cut} values confirms that pixels below an $\mathrm{S/N}$ of 12 yield poor fits or large uncertainties in critical parameters like radial velocities errors. The input parameters for one- and two-component fits are listed in Table~\ref{tab:guesses}.

\subsection{Output parameters and dependencies} \label{sec:output_par}
\texttt{PySpecKit} produces cubes containing the best-fit parameters (T$_{\mathrm{ex}}$, $\tau$, V$_{\mathrm{c}}$, and $\sigma$) and their associated uncertainties for each spectrum and velocity component. Using these parameters, we construct a modeled \nthp cube and separate the first (FVC) and second (SVC) velocity components into individual cubes. To evaluate the dependence of the fitting on the input parameters, we analyze the two most prominent \nthp filaments (R1 and R2; Fig.~\ref{fig:SPITZER_N2H+filaments}). We test the $start\_pixel$ parameter while keeping all other inputs fixed, performing 716 fits for R1 and 576 for R2, each using a different initial pixel. We find that the derived parameters and uncertainties are largely independent of the initial pixel. However, when fitting multiple velocity components, the starting pixel must correspond to a spectrum where the components are clearly resolved. We also test different parameter limits and find that our adopted ranges are sufficiently broad to avoid biasing the results.

\subsection{Output cleaning} \label{sec:cleaning}

With this modeling approach, some spectra may yield poor fits due to noise or an incorrect number of velocity components. Therefore, it is essential to identify the appropriate number of components and assess the quality of the fitted parameters and uncertainties. In particular, we analyze the error distributions of the fitted parameters. Errors in centroid velocity and velocity dispersion are mostly below 0.3\,km\,s$^{-1}$, with only 5 and 26 pixels exceeding this value, respectively. Since this threshold is conservative relative to the \nthp velocity resolution, we adopt it as the maximum allowed uncertainty for $\mathrm{V_c}$ and $\sigma$. We also find 184 fits with $\sigma$ uncertainties equal to 0\,km\,s$^{-1}$, indicating that the best-fit values lie at the edge of the allowed parameter range. These spectra likely contain unresolved velocity structure and may require multiple velocity components.

The parameter $\tau$ is more difficult to constrain because it can become arbitrarily large. We therefore apply a mask requiring $\tau/e(\tau) > 1$, where $e(\tau)$ is the uncertainty in $\tau$. Large $\tau$ values are typically associated with large uncertainties and flat-top spectral profiles, leading us to exclude $\sim$860 pixels from the model. Most retained $\tau$ values are below 25. For T$_{\mathrm{ex}}$, optically thin pixels ($\tau < 1$) often converge to the upper input limit with uncertainties of 0\,K. However, because T$_{\mathrm{ex}}$ has little impact on the spectral profile in this regime, we retain the original fitted values until the column density analysis (Sec.~\ref{sec:dens}).

Inspection of the rejected spectra shows that many are associated with multiple velocity components, including cases where the components are difficult to resolve. To address this, we refit the entire cube using two velocity components and evaluate whether the fits improve both inside and outside the previous selection criteria. For spectra containing a single velocity component, the two-component model often produces poor fits for either the bluest or reddest component. In these cases, we retain the single-component fit when its uncertainties are low and the fit quality is satisfactory. In addition, poor two-component fits are generally associated with velocity uncertainties larger than 0.3\,km\,s$^{-1}$ ($\sim$15$\%$ of pixels), large relative $\tau$ uncertainties ($\sim$30$\%$), zero-valued parameters (11$\%$ of pixels), and high $\tau$ values. We additionally impose an upper limit of $\tau = 100$ for each velocity component, removing 52 pixels above this threshold.

\begin{figure}[h!]
	\centering
    \includegraphics[scale = 0.3]{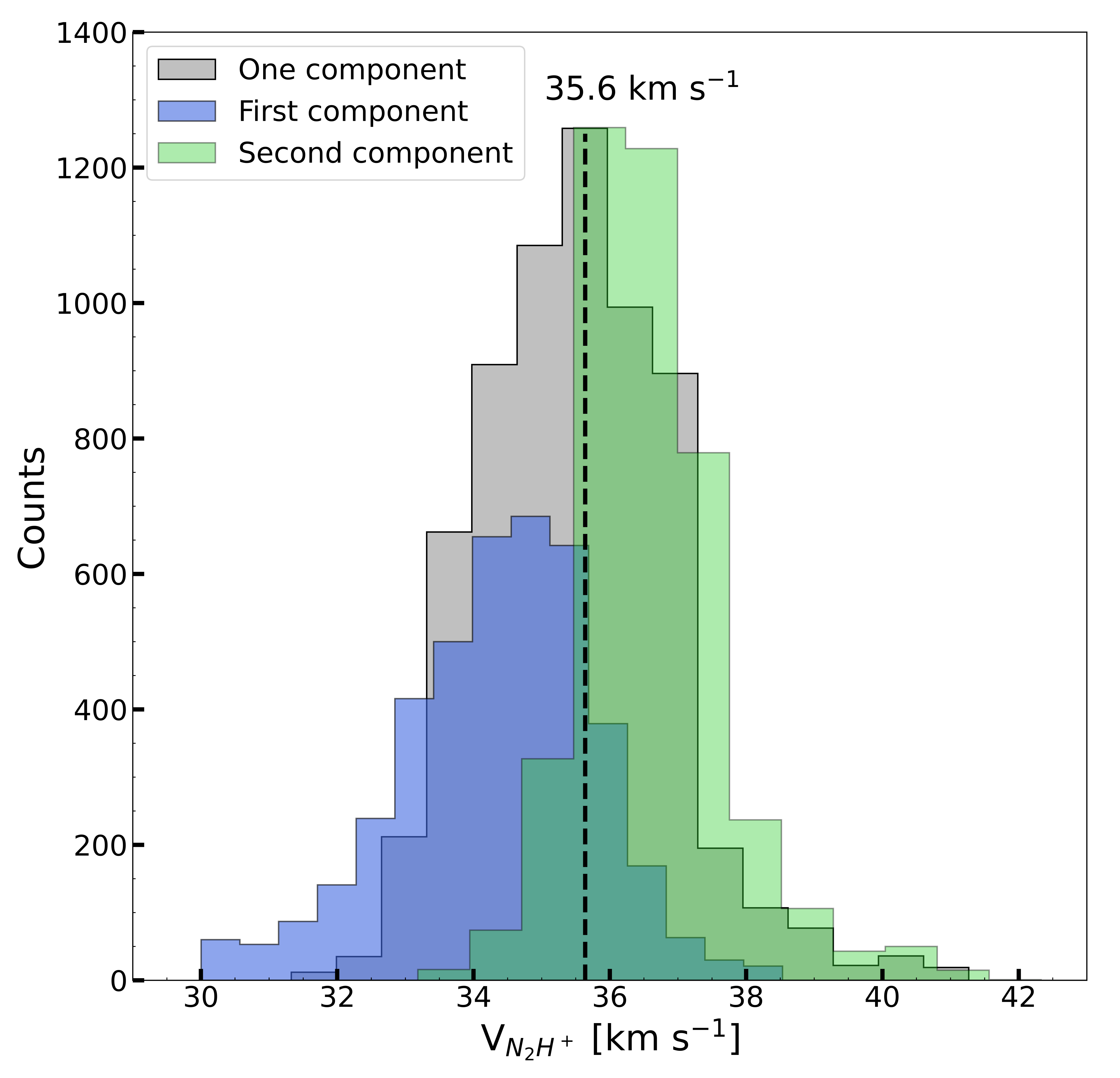}
    \caption{Histograms of the velocity distributions for single- (grey) and two-component (blue and green) velocity fits to the spectra (see text). The black dashed line shows the velocity boundary, at 35.6\,$\kms$, used to distribute one velocity component fits in the merged model cube.}
    \label{fig:histogram}
\end{figure}

In total, we obtain $\sim$\,15,000 spectra with a well defined model for one ($\sim$\,45$\%$) or two ($\sim$\,55$\%$) velocity components. Two velocity components are spatially distributed in the whole region, without a preferred location, but with a slight concentration in the R2 filament and in regions with large $\sigma$ values.

\subsection{Model cube merging}\label{sec:merge}

\begin{figure*}[h!]
  \centering
  \includegraphics[trim=0cm 0cm 0cm 0cm, clip,width=0.8\linewidth]{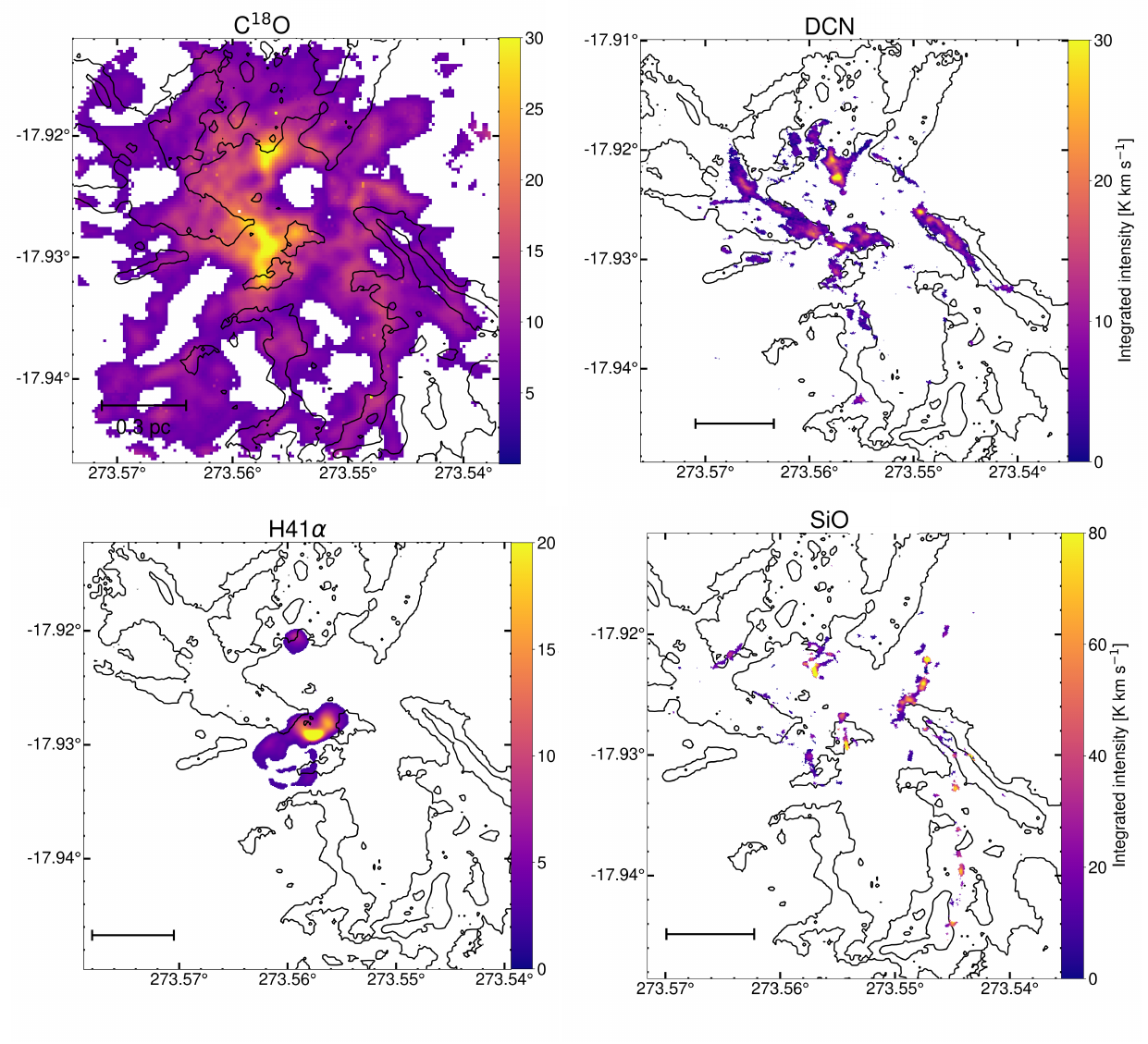}
  \vspace{-0.5cm}
  \caption{Integrated intensity of the complementary tracers: \ceto \citep[$J$=2-1, upper-left panel;][]{atanu2024}, SiO \citep[$J$=5-4, upper-right panel;][]{towner2024}, \hfoa \citep[bottom-left panel;][]{roberto2024}, and DCN \citep[$J$=3-2, bottom-right panel;][]{DCN}. The black contour shows \nthp integrated intensity at 25 and 100\,K\,$\kms$. The scale bar indicates 0.3\,pc at the distance of the G012 protocluster. The \hfoa recombination line and \ceto emission are mostly concentrated at the center of G012, where \nthp is mostly absent and an OB type stars cluster is located (see Fig.~\ref{fig:SPITZER_N2H+filaments}). In the areas surrounding the main \nthp filaments, we observe a distribution of lower \ceto integrated intensity emission ($\sim$\,15\,K\,km\,s$^{-1}$). DCN traces some regions of the main \nthp structures, and its emission is associated with cores and filaments. Specifically, DCN dominates the top of the R2 filament, central regions of the protocluster, at the edges of R1, and surroundings of the \hfoa bubbles.}
  \label{fig:TRACERS}
\end{figure*}

We obtain the final modeled cube by merging the fits with single and double velocity component. To carry out this step we follow a similar method than the one performed by \citet{nico2024}. 

To separate the two velocity components, we define a boundary using the midpoint between the upper limit of the first component (mean plus one standard deviation) and the lower limit of the second component (mean minus one standard deviation). Components below 35.6\,$\kms$ ($\sim$3,535 pixels) are assigned to the first velocity component (FVC), while those above ($\sim$2,984 pixels) are assigned to the second velocity component (SVC). This results in two velocity structures containing 7,675 pixels in the FVC and 7,124 pixels in the SVC.

\begin{equation}
	\mathrm{limit} = \frac{\langle \mathrm{V_{c,1}} \rangle + \mathrm{std}(\mathrm{V_{c,1}}) + \langle \mathrm{V_{c,2}} \rangle - \mathrm{std}(\mathrm{V_{c,2}})}{2} \rightarrow \mathrm{35.6 } \mathrm{km s}^{-1}.
\end{equation}
Here $\mathrm{V_{c,1}}$ and $\mathrm{V_{c,2}}$ correspond to the velocity center parameter of the first (bluest) and second (reddest) velocity component from the double component model, respectively. This approach ensures the limit reflects a balanced division between the two velocity distributions observed in the histogram in Figure~\ref{fig:histogram}, resulting in a velocity boundary of 35.6\,$\kms$.

\subsection{Complementary tracers line fitting} \label{sec:ap_otherstracers}

For tracers not previously modeled (DCN, \hfoa, and SiO), we apply a single-Gaussian fit using \texttt{PySpecKit}. The model includes three free parameters: amplitude (Amp), centroid velocity (V$_{\mathrm{c}}$), and velocity dispersion ($\sigma$), with initial guesses derived from the moment maps. Given the simplicity and low computational cost of the model, we do not constrain the parameter ranges. We additionally apply tracer-dependent $\mathrm{S/N}$ thresholds through the \textit{signal$\_$cut} parameter. The $\mathrm{S/N}$ is calculated from the peak intensity of each spectrum divided by its RMS noise, estimated from emission-free channels (Table~\ref{tab:SPECTRALINES}).

\section{Alternative filament detection using FilFinder} \label{sec:ap_filfinder}

Our current filament selection is based on the visual identification of prominent and kinematically well-separated filamentary structures. More specifically, we are interested in the filamentary structures where the most clear and significatively kinematic patterns are. However, the R1 and R2 selected regions could be part of more extended structures or be conected to other protocluster regions. To addres this, and in order to validate the robustness of our filament selection, we use the FilFinder algortihm to independently identify filamentary structures, which we then compare to our original selection.

\begin{table}[h!]
\centering
\caption{FilFinder parameters adopted for the filament detection.}
\label{tab:filfinder_params}
\begin{tabular}{l             l             l}
\hline
Parameter & Value (pixels) & Physical scale \\
\hline
smooth size$^{\mathrm{a}}$ & 5 pix & 0.05 pc \\
size thresh$^{\mathrm{b}}$ & 550 pix$^2$ & 0.05 pc$^2$ \\
adapt thresh$^{\mathrm{c}}$ & 50 pix & 0.49 pc  \\
fill hole size$^{\mathrm{d}}$ & 20 pix$^2$ & 0.002 pc$^2$  \\
skel thresh$^{\mathrm{e}}$ & 80 pix & 0.78 pc  \\
branch thresh$^{\mathrm{f}}$ & 20 pix & 0.20 pc \\ \hline

\end{tabular}\\
\footnotesize{
(a) Gaussian smoothing scale applied before the mask definition.\\
(b) Minimum area required for a filamentary structure to be retained. (c) Size of adaptive thresholding window that defines the spatial scale over which the local background intensity is estimated to determine whether a pixel belongs to a filamentary structure. (d) Maximum hole area filled within detected regions. (e) Minimum filament skeleton length retained. (f) Minimum branch length retained during pruning.}
\end{table}

\begin{figure*}[h!]
  \centering
  \includegraphics[trim=0cm 20cm 0cm 0cm, clip,width=0.9\linewidth]{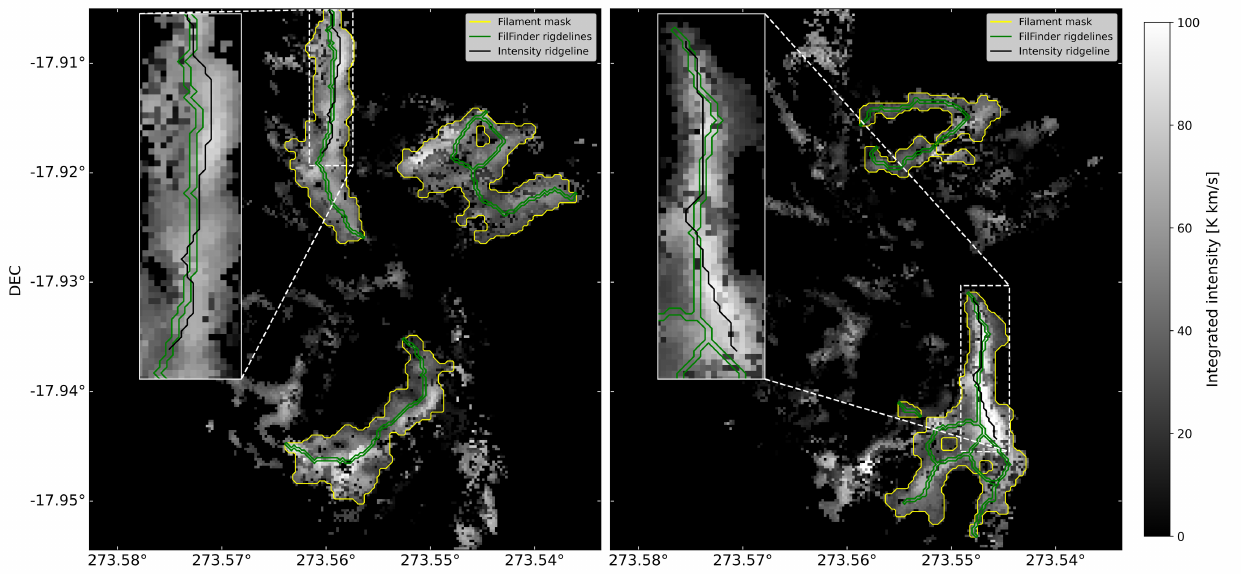} 
  \vspace{-0.3cm}
  \caption{Results from FilFinder filament detection over the FVC (left panel) and SVC (right panel) integrated intensity maps. In both panels: yellow contours represent the final masking area where we apply the algorithm (estimated by SNR and spatial extension considerations), green rigdelines represent the skeletons and branches identifying by FilFinder, black rigdelines shows the filament spine identification in Sec. \ref{sec:gradients} based on the integrated intensity emission in the R1 and R2 regions. We add a zoom in the R1 and R2 regions to compare the differences between FilFinder (green lines) and our previous (black lines) rigdeline identification.}
  \label{fig:filfinder_detection}
\end{figure*}

\subsection{Method}

We use the \texttt{FilFinder} python package \citep{filfinder} to identify and characterize filamentary structures in 2D emission maps. The analysis is performed on the integrated intensity maps of each velocity component, since the structures are primarily separated in velocity space (upper panels of Fig.~\ref{fig:structures}). Filament detection is restricted to regions with $\mathrm{S/N} > 12$. The images are first preprocessed using percentile-based flattening to suppress bright compact emission and enhance extended low-contrast structures. Filament masks are then generated through adaptive thresholding, combined with minimum area criteria and optional hole filling to remove small-scale noise features. The minimum filament sizes are selected based on the spatial resolution of the data. Table~\ref{tab:filfinder_params} summarizes the adopted mask parameters prior to filament identification. All spatial thresholds (see below) are larger than the beam minor axis (2.1" $\approx$ 0.024~pc), ensuring that the detected structures are spatially resolved filaments.

We adopt a smoothing scale of 5 pixels (0.05~pc; $\sim$2 beam sizes) to reduce pixel-scale noise while preserving resolved filamentary structures. A minimum area threshold of $\sim$0.05~pc$^2$ is used to reject compact or marginally resolved features. The adaptive threshold window is set to $\sim$0.5~pc, matching the characteristic filament scales. We retain only skeletons longer than 0.8~pc to focus on parsec-scale filaments, and prune branches shorter than 0.2~pc to remove minor substructures while preserving the main filamentary morphology. From the final mask, the algorithm computes a medial skeleton representation, reducing each filament to a one-pixel-wide spine (equidistant from the filament edges). This rigdeline estimation differ of our method to estimate the filament spine in Sec.~\ref{sec:gradients} where we use the integrated emission in each filament. The resulting skeleton network is analyzed to identify connected components and branches, and is pruned according to length thresholds in order to keep only the main filamentary paths.

\subsection{Variations in physical parameters}

In Fig.~\ref{fig:filfinder_detection} we present the results of the FilFinder detection applied to both velocity component integrated intensity maps (see upper panels in Fig.~\ref{fig:structures}). The final skeletons identified in the R1 and R2 regions appear more extended than assumed in our baseline definition. Nevertheless, in R1 we recover a ridgeline that closely follows the structure adopted in our analysis. In some cases, FilFinder identifies more complex morphologies, particularly in the R2 filament, where most of the elongated structure follows a single main skeleton, but the lower part splits into approximately four possible branches (see right panel in Fig.~\ref{fig:filfinder_detection}). Although one could select only the longest path as the main spine of the filament, we consider that it would not be fully reliable to claim that this represents the principal structure. Higher-resolution data will be required to confirm whether these features correspond to dependent or independent substructures near the filament end. For this reason, and to maintain consistency and simplicity in our main kinematic analysis, we restrict our study to the R1 and R2 regions defined at the beginning of this paper (see Fig.~\ref{fig:SPITZER_N2H+filaments}, right panel), which we consider our main filamentary structures of interest.

To quantify potential variations in our main analysis, we derived key parameters (mass estimation, mass profiles and associated metrics, SFR, and SFE) within the areas identified by FilFinder for the R1 and R2 regions. Within the yellow contours shown in Fig.~\ref{fig:filfinder_detection}, which trace the more extended filamentary areas, we derive an H$_2$ mass of 2450~M$_\odot$ for the extended R1 structure and 5000~M$_\odot$ for the extended R2 filament. These values are approximately a factor of two higher than those obtained using our baseline regions alone, emphasizing the significant contribution of these dominant filaments to the total protocluster mass. Overall, about 60$\%$ of the protocluster mass is contained within these main filamentary structures.

\begin{figure}[h!]
    \centering
    \includegraphics[scale = 0.3]{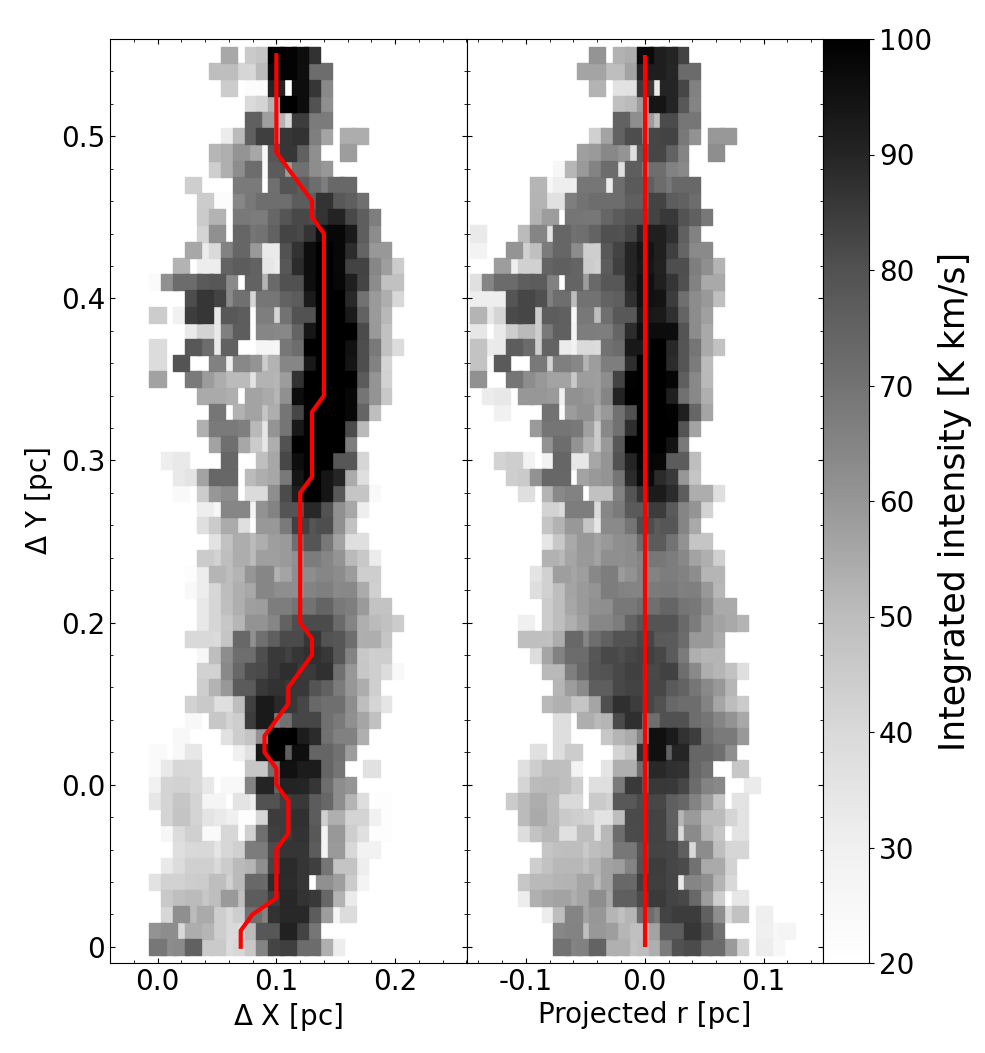}
    \vspace{-0.3cm}
    \caption{Example of the R1 filament alignment relative to the integrated intensity moment map. Left panel: Integrated intensity map of the FVC in the R1 region. The red line represent the ridgeline estimated based on the integrated intensity peak along to the filament. The X and Y axis represent the R1 length and width in units of pc. Right panel: Integrated intensity map aligned respect to the ridgeline. The X axis represent the projected radius result of the alignment process.}
\label{fig:rigdeline}
\end{figure}

The cumulative mass profiles of both extended filaments preserve the approximately linear behavior previously found for R1 and R2. We therefore evaluate how strongly the mass profiles change when considering the extended structures. To align the data, we adopt the FilFinder skeleton output as the ridgeline reference. For R1 we use the single skeleton returned by the algorithm, while for R2 we select the most extended skeleton as the filament ridgeline. On average, we find differences of about 30$\%$ relative to the metrics derived using only the R1 and R2 regions (see Table~\ref{tab:metrics} for details). The extended filaments show lower M/L values but larger projected radius, which is a direct consequence of the ridgeline alignment. Despite these lower M/L ratios compared to the baseline R1 and R2 regions, the profiles remain significantly higher than those measured in the comparison regions shown in Fig.~\ref{fig:enclosedmass}, reinforcing the conclusion that these filaments correspond to highly dense environments and are dominant mass reservoirs in the G012 protocluster.

In addition, we quantify the number of cores, their classification (prestellar or protostellar), and their masses to estimate the SFR and SFE within the extended areas. In the extended R1 filament, we now identify a protostellar contribution (in contrast to the original R1 selection, which contained only prestellar cores). In total, we detect nine cores in the R1 extension (two protostellar and seven prestellar) and 15 cores in the R2 extension (11 protostellar and 4 prestellar). Following the same timescale assumptions described in Sec.~\ref{sec:mass}, we estimate the SFR for each core population and for the total. For the extended R1 filament, we derive a prestellar SFR of 6.25 M$_{\odot}$~Myr$^{-1}$ and a protostellar SFR of 5.40 M$_{\odot}$~Myr$^{-1}$, resulting in a total SFR of 11.65 M$_{\odot}$~Myr$^{-1}$, corresponding to an increase of approximately 60$\%$ relative to the baseline region. For the extended R2 filament, we estimate a prestellar SFR of 6.3 M$_{\odot}$~Myr$^{-1}$ and a protostellar SFR of 47.5 M$_{\odot}$~Myr$^{-1}$, yielding a total SFR of 53.9 M$_{\odot}$~Myr$^{-1}$, which represents only a ~2$\%$ difference compared to the baseline region.

For the SFE, the extended R1 region has a total value of 0.0041 (with a prestellar contribution of 0.0030 and a protostellar contribution of 0.0011) that differ in around 60$\%$ with the SFE in the original R1 region, while the extended R2 region has a total SFE of 0.0062 (with a prestellar contribution of 0.0015 and a protostellar contribution of 0.0047), here the SFE differs only by 20$\%$ respect to our previous estimation in Sec. \ref{sec:mass}. This indicates that the R2 region forms cores with approximately twice the efficiency of the R1 filament. Despite the larger changes in SFR and SFE in the R1 filament, the extended R2 filament maintains a broader diversity in its core population, with a slight tendency toward more evolved cores, and a higher SFR than R1 (SFR in R2 is around a factor of five higher). The main difference compared to the parameters estimated within the smaller baseline regions is that the SFE gap between both filaments becomes smaller (SFE in R2 is around a factor of 1.5 higher than in R1).

The kinematic patterns of rotation and collapse discussed in the main analysis for R1 and R2 regions not change when considering the extended filaments. However, we highlight a kinematic transition between the upper and lower parts of the R1 filament. Specifically, the upper part exhibits the rotation signatures discussed in Sec.~\ref{sec:gradients}, whereas the lower part shows more dispersed velocity patterns and V-shaped structures that we discuss in Sec.\,\ref{sec:dis}. This transition may reflect the influence of the core population in that region or the impact of the feedback produced by the \hh bubbles in the region.

\section{Data alignment}\label{sec:ap_datalign}

We apply a filament alignment procedure throughout this work. For example, in Sec.~\ref{sec:gradients} the filaments are aligned relative to the peak integrated intensity to estimate average velocity gradients, while in Sec.~\ref{sec:linemass} the alignment is based on the column density distribution. Below we describe the procedure for the first case (Fig.~\ref{fig:rigdeline}), although the same method can be applied to other datasets:

\begin{enumerate}
	\item Map rotation: We first rotate the \nthp maps (e.g., integrated intensity and velocity centroid) to align the main filaments along the vertical axis (see Fig.~\ref{fig:pv_sub}, upper-left panel).
	
	\item Ridgeline estimation: Using the \nthp integrated intensity map, we estimate the filament ridgeline with a custom function that traces the peak intensity along the y-axis, defining the filament spine (red line in Fig.~\ref{fig:rigdeline}). The procedure is applied independently to the R1 and R2 filaments.
	
	\item Ridgeline smoothing: We smooth the ridgeline using a one-dimensional uniform filter from \texttt{scipy.ndimage} with a window size of 1 pixel. This reduces sharp fluctuations and produces a smoother representation of the filament path.
	
	\item Alignment and analysis: The maps are aligned to the smoothed ridgeline (Fig.~\ref{fig:rigdeline}, right panel), transforming the x-axis into projected radius and enabling a more robust analysis of velocity and mass profiles.
\end{enumerate}

\section{Technical considerations to estimate relative abundance} \label{sec:ap_relative}

Reprojection of the \nthp column density map can introduce edge artifacts because the regions with sufficient signal-to-noise are highly irregular. To mitigate this effect, we follow the methodology of \citet{nico2024} to identify affected pixels: a) We generate a binary mask from the \nthp column density map, assigning values of 1 to valid pixels and 0 to pixels below the signal-to-noise threshold. b) The mask is reprojected to the H$_2$ pixel scale of \citet{pierre2024}. Pixels with intermediate values between 0 and 1 identify reprojection artifacts, primarily affecting regions within 1-3 pixels of the image boundaries. c) We exclude all pixels with values <\,1 in the reprojected mask, producing a reprojection-artifact-free \nthp column density map.

To analyze the relative abundance distribution, we construct a histogram using the Freedman-Diaconis method \citep{nico2024}, obtaining an optimal bin width of 0.377\,$\times$\,10$^{-10}$ and 40 bins. We adopt the mode as the representative abundance value, since previous studies showed it is robust against changes in binning \citep{nico2024}. For the full sample, the mode is 0.84\,$\times$\,10$^{-10}$. To evaluate the impact of stricter $\tau$ uncertainty constraints on the abundance estimate, we apply progressively stronger masks: 

\begin{enumerate}[label=\alph*.]
	\item $\frac{\tau}{e(\tau)}$\,>\,1.5 ($\sim$\,3$\%$ pixels removed)
	\item $\frac{\tau}{e(\tau)}$\,>\,2 ($\sim$\,7$\%$ pixels removed)
	\item $\frac{\tau}{e(\tau)}$\,>\,3 ($\sim$\,15$\%$ pixels removed)
\end{enumerate}

The mode remains stable between 0.84\,$\times$\,10$^{-10}$ and 1.07\,$\times$\,10$^{-10}$. To minimize the impact of large $\tau$ uncertainties while preserving most of the filamentary structure, we adopt the $\tau/e(\tau) > 2$ mask, which removes only $\sim$7$\%$ of the pixels.

\begin{figure}[h!]
	\centering
    \includegraphics[scale = 0.32]{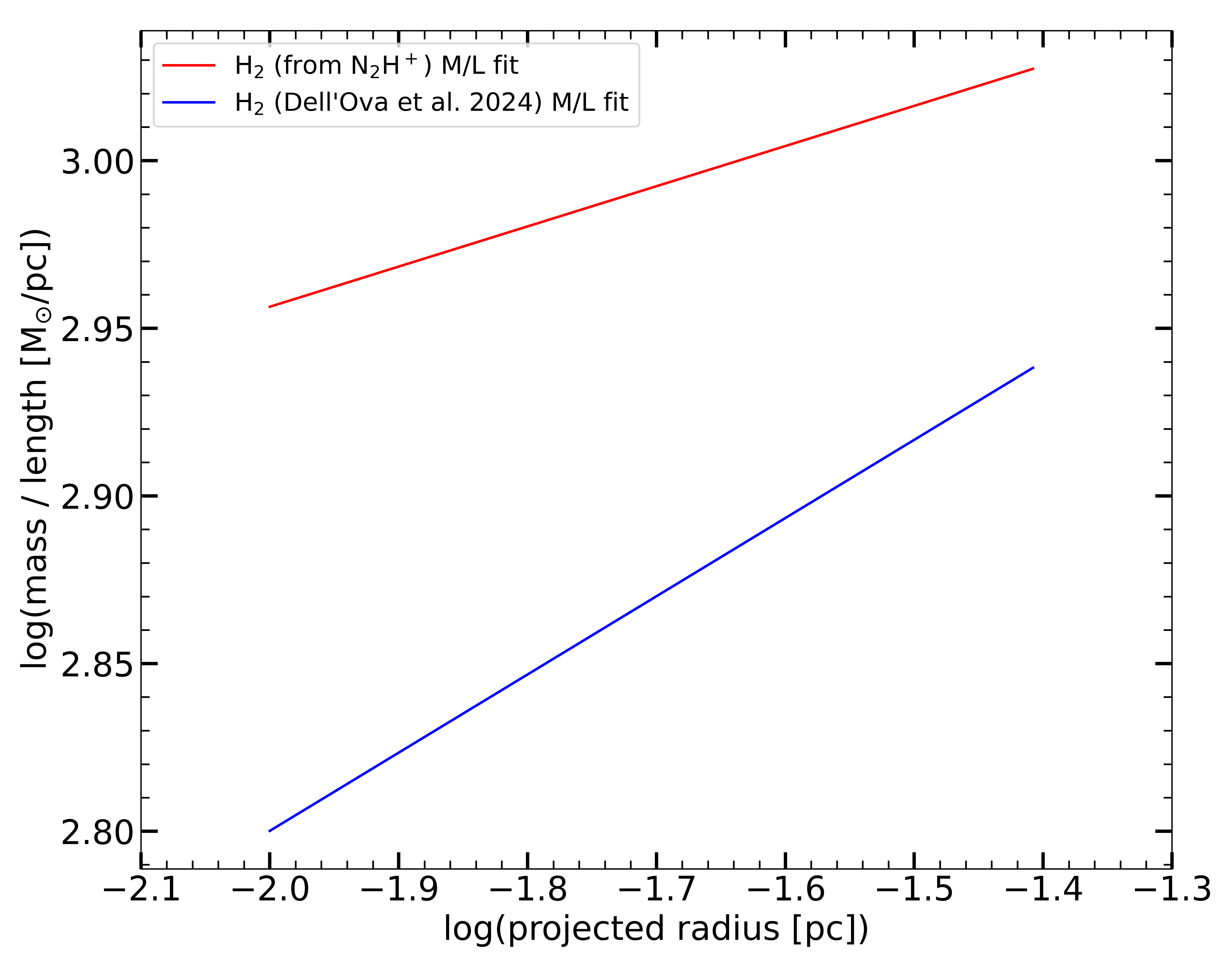}
    \vspace{-0.3cm}
    \caption{Mass over length profile of the selected region for the mass validation step. Red line represents the resulting profile considering H$_2$ mass estimated in Sec.~\ref{sec:mass}, while blue line shows the final profile considering H$_2$ mass from the column density estimated in \citet{pierre2024}. }
\label{fig:ap_enclosedmassval}
\end{figure}
\vspace{-0.5cm}

\section{Mass validation} \label{sec:ap_massval}

To validate the mass selection used for the line-mass profiles (Sec.~\ref{sec:linemass}), we compare the mass distributions derived from the H$_2$ column density map \citep{pierre2024} and the \nthp-based mass estimate from Sec.~\ref{sec:dens}. We select a well-detected elongated region within the R2 filament, with a length of 0.4\,pc and width of 0.03\,pc, and align the filament to compare their cumulative mass distributions. Both distributions follow a similar linear trend, differing only slightly in mass ($\Delta$M $\sim$ 300\,M$_{\odot}$). We therefore derive the M/L profiles following the method described in Sec.~\ref{sec:linemass}. The resulting profiles show minor differences in slope and the expected offset in normalization (Fig.~\ref{fig:ap_enclosedmassval}). We additionally estimate the normalization constants of the apparent volume density (Eq.~\ref{eq:volumedensity}), gravitational potential (Eq.~\ref{eq:gravitationalpot}), and gravitational acceleration (Eq.~\ref{eq:gravitationalacc}) using both mass distributions, finding average differences of $\sim$35$\%$.

\begin{figure}[h!]
    \centering
    \includegraphics[scale = 0.32]{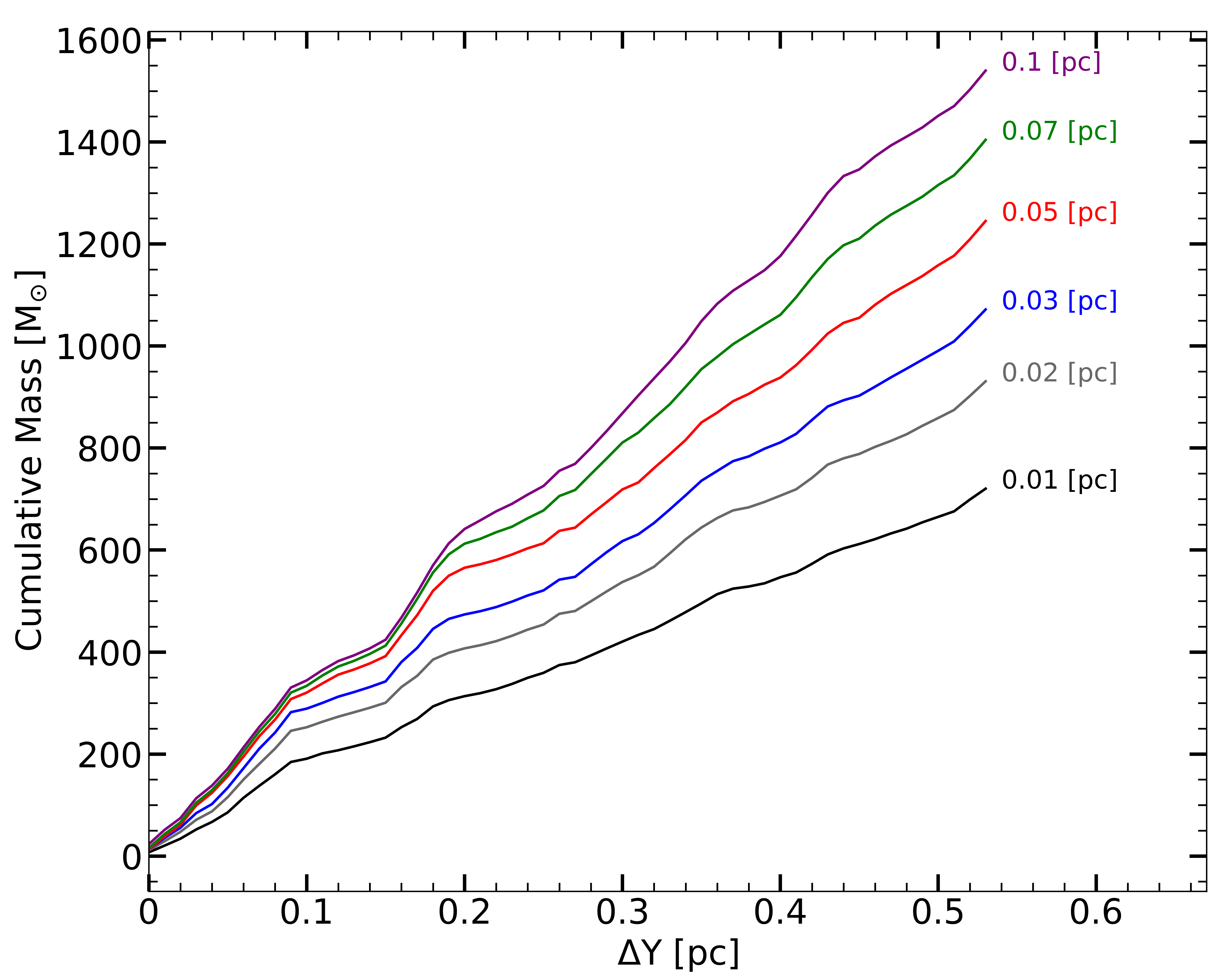}
    \vspace{-0.3cm}
    \caption{Example of cumulative mass profile in the R1 filament. Color curves highlights cumulative mass measurements at different filament radius. Profiles show a linear trend of the cumulative mass along the filament.}
\label{fig:R1cumulativemass}
\end{figure}

Given the overall agreement between both distributions and the improved mass profile obtained for the R1 filament, we adopt the total H$_2$ mass map (M$_{tot}$) derived from the \nthp data for the analysis in Sec.~\ref{sec:linemass}. Using M$_{tot}$, we obtain an approximately linear cumulative mass profile in the R1 filament over a filament length of 0.6\,pc (Fig.~\ref{fig:R1cumulativemass}).

\vspace{-0.5cm}

\section{Core velocities}\label{sec:cores}
We estimate the velocities of the G012 cores identified in the catalogs described in Sec.~\ref{sec:data} using \nthp and DCN spectral lines. Of the 38 cores previously detected in DCN \citep{DCN}, 36 are also detected in \nthp with $\mathrm{S/N} > 12$. We additionally use the 96 continuum cores identified by \citet{armante2024}, 58 of which lack previous velocity measurements. To derive velocities, we fit the DCN and \nthp emission with a single-Gaussian model (see Appendix~\ref{sec:ap_linefitting}). The major and minor axes of each core (Tables~\ref{tab:ap_ourcores} and \ref{tab:ap_dcncores}) define the region used to extract average spectra and estimate velocity centroids and mean fitted parameters.

From the \citet{armante2024} catalog, we identify 24 cores detected in both tracers (DCN $\cap$ \nthp sample), 20 detected only in \nthp, 2 detected only in DCN (DCN ``new'' sample), and 12 undetected in both tracers. We define V$_{\mathrm{DCN}}$ and V$_{\mathrm{N_2H^+}}$ as the velocity centroids derived from the DCN and \nthp fits, respectively. Since DCN traces denser, warmer, and more compact gas, we adopt V$_{\mathrm{DCN}}$ when available. To estimate velocities for the \nthp-only sample, we compute the offset V$_{\mathrm{DCN}}-$V$_{\mathrm{N_2H^+}}$ using 60 cores detected in both tracers: 36 from the previous DCN catalog and 24 from the DCN $\cap$ \nthp sample \citep[similar to the method applied in G353.41 by][]{yiyo2024}.

We find that \nthp velocities are blueshifted by $\sim$0.16\,$\kms$ relative to DCN, smaller than the spectral resolutions of both tracers (\nthp: 0.23\,$\kms$; DCN: 0.34\,$\kms$). Using this correction, we increase the sample of cores with velocity estimates by 48 objects: 24 from the DCN $\cap$ \nthp sample, 20 from the \nthp sample, and 2 from the DCN ``new'' sample.

\begin{table*}[h!]
    \begin{center}
    \caption{Cores velocities detected in this study.}
    \begin{tabular}{cccccccc}
        \multicolumn{8}{c}{\textit{DCN and \nthp detected sources}} \\\hline \hline
        ID$^{a}$ & RA & DEC & F(A) & F(B) & PA & Vel$^{b}$ & $\Delta$(V)$^{c}$ \\
         &  &  & [$^{\arcsec}$] & [$^{\arcsec}$] & [deg] & [km/s] & [km/s] \\ \hline 
        6 & 273.5557352 & -17.9151274 & 6.80 & 5.90 & 141.5 & 34.5 $\pm$ 0.030 & +0.08 \\ 
        15 & 273.5595282 & -17.9203648 & 14.7 & 11.7 & 209.6 & 34.5 $\pm$ 0.064 & +0.81 \\
        20 & 273.5585587 & -17.9208742 & 25.9 & 19.9 & 138.5 & 35.2 $\pm$ 0.029 & +0.29 \\
        21 & 273.5581258 & -17.9329013 & 12.2 & 10.9 & 217.1 & 36.2 $\pm$ 0.015 & -1.60 \\ 
        24 & 273.5420287 & -17.9327274 & 10.5 & 6.60 & 146.6 & 36.8 $\pm$ 0.064 & +0.27 \\ 
        27 & 273.5587301 & -17.9276784 & 9.10 & 6.50 & 172.6 & 34.8 $\pm$ 0.098 & -0.18 \\ 
        28 & 273.5574005 & -17.9281480 & 7.90 & 4.90 & 147.4 & 35.5 $\pm$ 0.068 & -0.53 \\ 
        35 & 273.5621830 & -17.9302229 & 23.7 & 15.1 & 211.8 & 35.5 $\pm$ 0.005 & +0.31 \\ 
        38 & 273.5511506 & -17.9252426 & 12.5 & 7.00 & 128.8 & 34.5 $\pm$ 0.001 & -1.42 \\ 
        45 & 273.5614832 & -17.9294106 & 12.0 & 8.40 & 207.1 & 34.8 $\pm$ 0.050 & -0.06 \\ 
        49 & 273.5512430 & -17.9221040 & 7.60 & 4.80 & 131.6 & 34.5 $\pm$ 0.022 & -1.35 \\ 
        51 & 273.5456695 & -17.9286524 & 7.40 & 6.90 & 144.9 & 37.5 $\pm$ 0.029 & +0.29 \\ 
        53 & 273.5559769 & -17.9272312 & 8.40 & 5.70 & 171.6 & 35.8 $\pm$ 0.039 & -4.85 \\
        55 & 273.5585399 & -17.9322517 & 9.00 & 8.80 & 233.3 & 35.8 $\pm$ 0.087 & -2.78 \\ 
        63 & 273.5559545 & -17.9280031 & 8.50 & 7.70 & 175.6 & 34.8 $\pm$ 0.081 & -1.01 \\
        72 & 273.5559098 & -17.9210339 & 11.3 & 7.50 & 186.2 & 34.5 $\pm$ 0.055 & -0.85 \\ 
        82 & 273.5541355 & -17.9372087 & 14.1 & 12.0 & 173.6 & 37.2 $\pm$ 0.014 & +0.94 \\ 
        84 & 273.5466969 & -17.9270489 & 6.40 & 4.80 & 149.7 & 38.5 $\pm$ 0.076 & +0.61 \\ 
        87 & 273.5661498 & -17.9243080 & 7.90 & 6.40 & 140.5 & 35.2 $\pm$ 0.017 & +0.09 \\ 
        92 & 273.5644636 & -17.9275000 & 11.4 & 9.20 & 140.4 & 35.5 $\pm$ 0.029 & +0.47 \\ 
        94 & 273.5573333 & -17.9209924 & 9.30 & 5.80 & 165.6 & 33.1 $\pm$ 0.036 & -1.65 \\ 
        96 & 273.5538282 & -17.9203791 & 15.8 & 11.2 & 155.9 & 36.5 $\pm$ 0.056 & +0.24 \\ 
        97 & 273.5534924 & -17.9302818 & 7.20 & 5.00 & 154.1 & 36.5 $\pm$ 0.054 & +0.71 \\ 
        101 & 273.5578973 & -17.9200468 & 10.5 & 8.00 & 93.50 & 35.8 $\pm$ 0.035 & +0.15 \\
         &  &  &  &  &  &  &    \\
        \multicolumn{8}{c}{\textit{\nthp detected sources}} \\ \hline \hline
        4 & 273.5481868 & -17.9458387 & 7.50 & 5.00 & 179.0 & 35.2$^{*}$ $\pm$ 0.009 & --- \\ 
        9 & 273.5541976 & -17.9141605 & 7.40 & 4.40 & 166.3 & 33.2 $\pm$ 0.010 & --- \\ 
        30 & 273.5472394 & -17.9220959 & 10.0 & 9.00 & 178.6 & 36.4 $\pm$ 0.075 & --- \\ 
        31 & 273.5714747 & -17.9249529 & 6.30 & 4.10 & 159.5 & 35.7 $\pm$ 0.005 & --- \\ 
        47 & 273.5573736 & -17.9253241 & 7.70 & 5.90 & 169.2 & 36.8 $\pm$ 0.047 & --- \\ 
        48 & 273.5599387 & -17.9247874 & 11.2 & 7.80 & 133.0 & 37.8 $\pm$ 0.066 & --- \\ 
        54 & 273.5741063 & -17.9183977 & 6.40 & 5.00 & 147.4 & 36.1 $\pm$ 0.012 & --- \\ 
        66 & 273.5574878 & -17.9319110 & 9.30 & 8.30 & 246.9 & 38.7 $\pm$ 0.021 & --- \\ 
        68 & 273.5655452 & -17.9186858 & 14.4 & 12.7 & 235.1 & 36.4 $\pm$ 0.040 & --- \\ 
        70 & 273.5425468 & -17.9318786 & 10.4 & 7.20 & 254.2 & 37.1 $\pm$ 0.006 & --- \\ 
        71 & 273.5596700 & -17.9235735 & 10.8 & 6.60 & 176.5 & 38.2 $\pm$ 0.055 & --- \\ 
        75 & 273.5384638 & -17.9342166 & 6.20 & 5.00 & 140.1 & 36.8 $\pm$ 0.009 & --- \\ 
        76 & 273.5562608 & -17.9398390 & 21.8 & 18.2 & 184.3 & 35.7 $\pm$ 0.015 & --- \\ 
        77 & 273.5402299 & -17.9333798 & 8.70 & 7.10 & 121.0 & 36.6 $\pm$ 0.006 & --- \\ 
        81 & 273.5453666 & -17.9406316 & 7.10 & 6.00 & 246.7 & 35.7 $\pm$ 0.014 & --- \\ 
        83 & 273.5609797 & -17.9336253 & 10.5 & 7.00 & 182.6 & 35.5 $\pm$ 0.056 & --- \\ 
        86 & 273.5612818 & -17.9299304 & 9.30 & 6.60 & 192.0 & 34.8 $\pm$ 0.049 & --- \\ 
        88 & 273.5693683 & -17.9265363 & 20.1 & 14.8 & 97.12 & 34.3 $\pm$ 0.031 & --- \\ 
        90 & 273.5455200 & -17.9412705 & 9.40 & 7.20 & 202.1 & 35.2 $\pm$ 0.013 & --- \\ 
        95 & 273.5437775 & -17.9373519 & 7.50 & 6.10 & 239.4 & 36.1 $\pm$ 0.007 & --- \\ 
         &  &  &  &   &  &  &    \\
        \multicolumn{8}{c}{\textit{DCN detected sources}} \\ \hline \hline
        79 & 273.5576220 & -17.9229346 & 8.90 & 5.80 & 179.1 & 37.2 $\pm$ 0.030 & --- \\
        100 & 273.5529552 & -17.9256818 & 7.60 & 4.60 & 105.0 & 35.8 $\pm$ 0.030 & --- \\ \hline
    \end{tabular}
    \label{tab:ap_ourcores}
    \end{center}
    \footnotesize{(a) We conserve the core ID from \citet{armante2024} for reference. (b) Centroid velocity estimated through DCN line-fitting (see Sec.~\ref{sec:cores}). (c) V$_{\text{DCN}}$ - V$_{\text{N}_2\text{H}^+}$, positive and negative symbols indicate if the \nthp velocity is blueshifted or redshifted relative to the DCN core velocities. (*) For cores without DCN emission we used \nthp velocity center (and associated uncertainties) extracted from the \texttt{PySpecKit} line-fitting. }

\end{table*}

\begin{table*}[h!]
\begin{center}
    \caption{\nthp velocity differences for DCN core catalog by \citet{DCN}}
    \vspace{-0.5cm}
    \begin{tabular}{cccccccc} \\\hline \hline 
        ID & RA & DEC & F(A) & F(B) & PA & Vel$^{a}$ & $\Delta$(V)$^{b}$ \\ 
         &  &  & [$^{\arcsec}$] & [$^{\arcsec}$] & [deg] & [km/s] & [km/s] \\ \hline 
        1.0 & 273.5493292 & -17.9256817 & 1.60 & 1.17 & 57.00 & 37.06 $\pm$ 0.06 & +1.36 \\ 
        3.0 & 273.5573504 & -17.9225106 & 1.84 & 1.52 & 8.000 & 36.16 $\pm$ 0.04 & -0.24 \\ 
        4.0 & 273.5444266 & -17.9375329 & 1.41 & 1.09 & 91.00 & 36.16 $\pm$ 0.08 & +0.66 \\ 
        6.0 & 273.5531484 & -17.9208183 & 1.59 & 1.11 & 54.00 & 32.60 $\pm$ 0.12 & -3.80 \\ 
        8.0 & 273.5486150 & -17.9262103 & 2.35 & 1.54 & 69.00 & 36.50 $\pm$ 0.03 & +0.10 \\ 
        11.0 & 273.5547682 & -17.9278997 & 1.72 & 1.42 & 76.00 & 35.82 $\pm$ 0.03 & +0.32 \\ 
        12.0 & 273.5444858 & -17.9302895 & 1.50 & 1.37 & 71.00 & 36.87 $\pm$ 0.04 & +0.07 \\ 
        13.0 & 273.5484365 & -17.9412478 & 1.75 & 1.54 & 17.00 & 35.40 $\pm$ 0.05 & +0.60 \\ 
        16.0 & 273.5526754 & -17.9285675 & 1.35 & 1.07 & 66.00 & 33.87 $\pm$ 0.03 & -0.43 \\ 
        19.0 & 273.5561582 & -17.9212927 & 1.78 & 1.46 & 176.0 & 34.78 $\pm$ 0.06 & -0.92 \\ 
        20.0 & 273.5569282 & -17.9233073 & 1.82 & 1.27 & 40.00 & 36.27 $\pm$ 0.08 & --- \\ 
        22.0 & 273.5661028 & -17.9234199 & 1.56 & 1.18 & 54.00 & 35.44 $\pm$ 0.02 & +0.64 \\ 
        23.0 & 273.5464417 & -17.9286205 & 1.68 & 1.35 & 61.00 & 37.05 $\pm$ 0.02 & +0.25 \\ 
        24.0 & 273.5689656 & -17.9390182 & 2.10 & 1.31 & 67.00 & 37.31 $\pm$ 0.08 & +0.51 \\ 
        25.0 & 273.5690243 & -17.9247231 & 1.85 & 1.23 & 90.00 & 36.09 $\pm$ 0.03 & +0.39 \\ 
        27.0 & 273.5561592 & -17.9348499 & 1.97 & 1.82 & 171.0 & 37.53 $\pm$ 0.02 & +0.73 \\ 
        29.0 & 273.5516652 & -17.9189507 & 1.91 & 1.67 & 112.0 & 35.87 $\pm$ 0.09 & +0.37 \\ 
        30.0 & 273.5576077 & -17.9311763 & 1.68 & 1.49 & 15.00 & 40.00 $\pm$ 0.07 & +0.90 \\ 
        31.0 & 273.5676865 & -17.9204887 & 2.18 & 1.55 & 158.0 & 33.52 $\pm$ 0.07 & -5.58 \\ 
        32.0 & 273.5470383 & -17.9333909 & 1.60 & 1.32 & 29.00 & 35.39 $\pm$ 0.08 & +0.59 \\ 
        33.0 & 273.5581592 & -17.9252346 & 2.16 & 1.68 & 25.00 & 35.31 $\pm$ 0.02 & -1.79 \\ 
        34.0 & 273.5465089 & -17.9276804 & 1.63 & 1.42 & 37.00 & 37.57 $\pm$ 0.02 & +0.47 \\ 
        36.0 & 273.5569394 & -17.9239001 & 1.71 & 1.45 & 67.00 & 35.82 $\pm$ 0.03 & --- \\ 
        39.0 & 273.5665414 & -17.9228278 & 1.80 & 1.36 & 49.00 & 35.17 $\pm$ 0.02 & +0.67 \\ 
        40.0 & 273.5457926 & -17.9284661 & 2.12 & 1.76 & 40.00 & 37.31 $\pm$ 0.02 & +0.21 \\ 
        41.0 & 273.5662283 & -17.9256745 & 1.69 & 1.47 & 48.00 & 35.02 $\pm$ 0.07 & +0.02\\ 
        42.0 & 273.5648070 & -17.9262439 & 2.01 & 1.65 & 97.00 & 36.09 $\pm$ 0.06 & +1.79 \\ 
        45.0 & 273.5557839 & -17.9182388 & 2.75 & 2.44 & 7.000 & 35.99 $\pm$ 0.08 & +0.49 \\ 
        46.0 & 273.5579243 & -17.9306652 & 1.83 & 1.75 & 44.00 & 39.71 $\pm$ 0.05 & +1.71 \\ 
        47.0 & 273.5479984 & -17.9272905 & 2.22 & 1.72 & 43.00 & 35.98 $\pm$ 0.02 & +0.08 \\ 
        51.0 & 273.5458037 & -17.9289719 & 1.75 & 1.43 & 77.00 & 37.23 $\pm$ 0.02 & +0.43 \\ 
        52.0 & 273.5503813 & -17.9228494 & 1.60 & 1.12 & 67.00 & 34.41 $\pm$ 0.05 & -0.09 \\ 
        53.0 & 273.5463654 & -17.9421570 & 1.84 & 1.37 & 33.00 & 36.48 $\pm$ 0.07 & +1.28 \\ 
        57.0 & 273.5614471 & -17.9187094 & 2.34 & 1.89 & 61.00 & 35.71 $\pm$ 0.02 & -0.39 \\ 
        58.0 & 273.5595440 & -17.9178954 & 1.59 & 1.45 & 158.0 & 34.27 $\pm$ 0.07 & -0.23 \\ 
        61.0 & 273.5475497 & -17.9266400 & 1.69 & 1.04 & 59.00 & 37.80 $\pm$ 0.02 & +0.70 \\ 
        62.0 & 273.5533917 & -17.9280329 & 2.13 & 1.63 & 94.00 & 34.03 $\pm$ 0.02 & -0.27 \\ 
        65.0 & 273.5597707 & -17.9186860 & 1.70 & 1.52 & 51.00 & 34.85 $\pm$ 0.06 & -0.35 \\ \hline
    \end{tabular}
    \label{tab:ap_dcncores}
    \end{center}
    \footnotesize{(a) DCN V$_{\mathrm{LSR}}$ (and associated uncertainties) extracted from \citet{DCN}. (b) V$_{\text{DCN}}$ - V$_{\text{N}_2\text{H}^+}$, positive and negative symbols indicate if the \nthp velocity is blueshifted or redshifted relative to the DCN core velocities.}
    
\end{table*}

\end{document}